
\input amssym.tex
\input epsf

\def\e{{\rm e}}

\def\ppal{\hbox{\accent23{$\!\!\!\! p$}}}

\def\bppiu{{\bf p}^{\scriptscriptstyle {+}}}

\def\lrD{\stackrel{\leftrightarrow}{D}}
\def\rhobar{{\bar \rho}}

\def\cO{{\cal O}}

\def\lrD{{\mathop{D}\limits^\leftrightarrow_{}}\ }

\def\hi{\hat {\imath}}
\def\hmu{\hat \mu}

\def\hk{\hat k}

\def\bl{{\bf l}}
\def\bm{{\bf m}}
\def\bn{{\bf n}}

\def\bp{{\bf p}}
\def\bq{{\bf q}}

\def\bu{{\bf u}}
\def\bv{{\bf v}}

\def\bx{{\bf x}}
\def\by{{\bf y}}
\def\bz{{\bf z}}

\def\bzero{{\bf 0}}


\magnification=\magstephalf
\hsize=16.0 true cm
\vsize=21.0 true cm
\hoffset=0.0 true cm
\voffset=1.0 true cm

\abovedisplayskip=12pt plus 3pt minus 3pt
\belowdisplayskip=12pt plus 3pt minus 3pt
\parindent=0.0em


\font\sixrm=cmr6
\font\eightrm=cmr8
\font\ninerm=cmr9

\font\sixi=cmmi6
\font\eighti=cmmi8
\font\ninei=cmmi9

\font\sixsy=cmsy6
\font\eightsy=cmsy8
\font\ninesy=cmsy9

\font\sixbf=cmbx6
\font\eightbf=cmbx8
\font\ninebf=cmbx9

\font\eightit=cmti8
\font\nineit=cmti9

\font\eightsl=cmsl8
\font\ninesl=cmsl9

\font\sixss=cmss8 at 8 true pt
\font\sevenss=cmss9 at 9 true pt
\font\eightss=cmss8
\font\niness=cmss9
\font\tenss=cmss10

 at 12 true pt
\font\bigrm=cmr10 at 12 true pt
 at 12 true pt

\font\Bigbss=cmssbx12 at 16 true pt
 at 14 true pt
 at 14 true pt
 at 16 true pt
 at 14 true pt

\font\tencmmib=cmmib10 \skewchar\tencmmib='177
\newfam\cmmibfam
\textfont\cmmibfam=\tencmmib

\catcode`@=11
\newfam\ssfam

\def\tenpoint{\def\rm{\fam0\tenrm}%
    \textfont0=\tenrm \scriptfont0=\sevenrm \scriptscriptfont0=\fiverm
    \textfont1=\teni  \scriptfont1=\seveni  \scriptscriptfont1=\fivei
    \textfont2=\tensy \scriptfont2=\sevensy \scriptscriptfont2=\fivesy
    \textfont3=\tenex \scriptfont3=\tenex   \scriptscriptfont3=\tenex
    \textfont\itfam=\tenit                  \def\it{\fam\itfam\tenit}%
    \textfont\slfam=\tensl                  \def\sl{\fam\slfam\tensl}%
    \textfont\bffam=\tenbf \scriptfont\bffam=\sevenbf
    \scriptscriptfont\bffam=\fivebf
                                            \def\bf{\fam\bffam\tenbf}%
    \textfont\ssfam=\tenss \scriptfont\ssfam=\sevenss
    \scriptscriptfont\ssfam=\sevenss
                                            \def\ss{\fam\ssfam\tenss}%
    \normalbaselineskip=19pt
    \setbox\strutbox=\hbox{\vrule height8.5pt depth3.5pt width0pt}%
    \let\big=\tenbig
    \normalbaselines\rm}

\def\ninepoint{\def\rm{\fam0\ninerm}%
    \textfont0=\ninerm      \scriptfont0=\sixrm
                            \scriptscriptfont0=\fiverm
    \textfont1=\ninei       \scriptfont1=\sixi
                            \scriptscriptfont1=\fivei
    \textfont2=\ninesy      \scriptfont2=\sixsy
                            \scriptscriptfont2=\fivesy
    \textfont3=\tenex       \scriptfont3=\tenex
                            \scriptscriptfont3=\tenex
    \textfont\itfam=\nineit \def\it{\fam\itfam\nineit}%
    \textfont\slfam=\ninesl \def\sl{\fam\slfam\ninesl}%
    \textfont\bffam=\ninebf \scriptfont\bffam=\sixbf
                            \scriptscriptfont\bffam=\fivebf
                            \def\bf{\fam\bffam\ninebf}%
    \textfont\ssfam=\niness \scriptfont\ssfam=\sixss
                            \scriptscriptfont\ssfam=\sixss
                            \def\ss{\fam\ssfam\niness}%
    \normalbaselineskip=16pt
    \setbox\strutbox=\hbox{\vrule height8.0pt depth3.0pt width0pt}%
    \let\big=\ninebig
    \normalbaselines\rm}

\def\eightpoint{\def\rm{\fam0\eightrm}%
    \textfont0=\eightrm      \scriptfont0=\sixrm
                             \scriptscriptfont0=\fiverm
    \textfont1=\eighti       \scriptfont1=\sixi
                             \scriptscriptfont1=\fivei
    \textfont2=\eightsy      \scriptfont2=\sixsy
                             \scriptscriptfont2=\fivesy
    \textfont3=\tenex        \scriptfont3=\tenex
                             \scriptscriptfont3=\tenex
    \textfont\itfam=\eightit \def\it{\fam\itfam\eightit}%
    \textfont\slfam=\eightsl \def\sl{\fam\slfam\eightsl}%
    \textfont\bffam=\eightbf \scriptfont\bffam=\sixbf
                             \scriptscriptfont\bffam=\fivebf
                             \def\bf{\fam\bffam\eightbf}%
    \textfont\ssfam=\eightss \scriptfont\ssfam=\sixss
                             \scriptscriptfont\ssfam=\sixss
                             \def\ss{\fam\ssfam\eightss}%
    \normalbaselineskip=10pt
    \setbox\strutbox=\hbox{\vrule height7.0pt depth2.0pt width0pt}%
    \let\big=\eightbig
    \normalbaselines\rm}

\def\tenbig#1{{\hbox{$\left#1\vbox to8.5pt{}\right.\n@space$}}}
\def\ninebig#1{{\hbox{$\textfont0=\tenrm\textfont2=\tensy
                       \left#1\vbox to7.25pt{}\right.\n@space$}}}
\def\eightbig#1{{\hbox{$\textfont0=\ninerm\textfont2=\ninesy
                       \left#1\vbox to6.5pt{}\right.\n@space$}}}

\font\sectionfont=cmbx10
\font\subsectionfont=cmti10

\def\figurecaptionfont{\ninepoint}
\def\tablecaptionfont{\ninepoint}


\newcount\equationno
\newcount\bibitemno
\newcount\figureno
\newcount\tableno

\equationno=0
\bibitemno=0
\figureno=0
\tableno=0


\footline={\ifnum\pageno=0{\hfil}\else
{\hss\rm\the\pageno\hss}\fi}


\def\section #1. #2 \par
{\vskip0pt plus .20\vsize\penalty-100 \vskip0pt plus-.20\vsize
\vskip 1.4 true cm plus 0.2 true cm minus 0.2 true cm
\global\def\equationlabel{#1}
\global\equationno=0
\leftline{\sectionfont #1. #2}\par
\immediate\write\terminal{Section #1. #2}
\vskip 0.4 true cm plus 0.1 true cm minus 0.1 true cm
\noindent}


\def\subsection #1 \par
{\vskip0pt plus 0.4 true cm\penalty-50 \vskip0pt plus-0.4 true cm
\vskip2.5ex plus 0.1ex minus 0.1ex
\leftline{\subsectionfont #1}\par
\immediate\write\terminal{Subsection #1}
\vskip1.0ex plus 0.1ex minus 0.1ex
\noindent}


\def\appendix #1 #2 \par
{\vskip0pt plus .20\vsize\penalty-100 \vskip0pt plus-.20\vsize
\vskip 1.6 true cm plus 0.2 true cm minus 0.2 true cm
\global\def\equationlabel{\hbox{\rm#1}}
\global\equationno=0
\leftline{\sectionfont Appendix #1 \quad {\it #2}}\par
\immediate\write\terminal{Appendix #1 #2}
\vskip 0.7 true cm plus 0.1 true cm minus 0.1 true cm
\noindent}



\def\equation#1{$$\displaylines{\qquad #1}$$}
\def\enum{\global\advance\equationno by 1
\hfill\llap{(\equationlabel.\the\equationno)}}


\def\ifundefined#1{\expandafter\ifx\csname#1\endcsname\relax}

\def\ref#1{\ifundefined{#1}?\immediate\write\terminal{unknown reference
on page \the\pageno}\else\csname#1\endcsname\fi}

\newwrite\terminal
\newwrite\bibitemlist

\def\bibitem#1#2\par{\global\advance\bibitemno by 1
\immediate\write\bibitemlist{\string\def
\expandafter\string\csname#1\endcsname
{\the\bibitemno}}
\item{[\the\bibitemno]}#2\par}

\def\beginbibliography{
\vskip0pt plus .15\vsize\penalty-100 \vskip0pt plus-.15\vsize
\vskip 1.2 true cm plus 0.2 true cm minus 0.2 true cm
\leftline{\sectionfont References}\par
\immediate\write\terminal{References}
\immediate\openout\bibitemlist=biblist
\frenchspacing\parindent=1.8em
\vskip 0.5 true cm plus 0.1 true cm minus 0.1 true cm}

\def\endbibliography{
\immediate\closeout\bibitemlist
\nonfrenchspacing\parindent=1.0em}

%
%
\def\figurecaption#1{\global\advance\figureno by 1
\narrower\figurecaptionfont
Fig.~\the\figureno. #1}
\def\tablecaption#1{\global\advance\tableno by 1
\vbox to 0.5 true cm { }
\centerline{\tablecaptionfont%
Table~\the\tableno. #1}
\vskip-0.4 true cm}
\def\thicktablerule{\hrule height1pt}
\def\thintablerule{\hrule height0.4pt}
\tenpoint

\immediate\openin\bibitemlist=biblist
\ifeof\bibitemlist\immediate\closein\bibitemlist
\else\immediate\closein\bibitemlist
\input biblist \fi
%
%
\def\Source{{\Bbb S}}
\def\projector{{\Bbb P}}
\def\su3id{{\Bbb I}}
\def\Zset{{\Bbb Z}}
\def\Nset{{\Bbb N}}
%
%
\def\lvec#1{\setbox0=\hbox{$#1$}
    \setbox1=\hbox{$\scriptstyle\leftarrow$}
    #1\kern-\wd0\smash{
    \raise\ht0\hbox{$\raise1pt\hbox{$\scriptstyle\leftarrow$}$}}
    \kern-\wd1\kern\wd0}
\def\rvec#1{\setbox0=\hbox{$#1$}
    \setbox1=\hbox{$\scriptstyle\rightarrow$}
    #1\kern-\wd0\smash{
    \raise\ht0\hbox{$\raise1pt\hbox{$\scriptstyle\rightarrow$}$}}
    \kern-\wd1\kern\wd0}
%
%
\def\rD{\rvec{D}}
\def\lD{\lvec{D}}
%
%
\def\psibar{\overline{\psi}}

%
%
\def\tr{{\rm tr}}
\def\btheta{\hbox{$\mathchar"0\hexnumber@\cmmibfam12$}}
%
%
\def\NPB #1 #2 #3 {Nucl.~Phys.~{\bf#1} (#2)\ #3}
\def\NPBproc #1 #2 #3 {Nucl.~Phys.~B (Proc. Suppl.) {\bf#1} (#2)\ #3}
\def\PRD #1 #2 #3 {Phys.~Rev.~{\bf#1} (#2)\ #3}
\def\PLB #1 #2 #3 {Phys.~Lett.~{\bf#1} (#2)\ #3}
\def\PRL #1 #2 #3 {Phys.~Rev.~Lett.~{\bf#1} (#2)\ #3}
\def\PR  #1 #2 #3 {Phys.~Rep.~{\bf#1} (#2)\ #3}
\def\etal{{\it et al.}}

\pageno=0
\hfill ROM2F/2002/03
\vskip 1.3 true cm 
\centerline
{\Bigbss   Moments of singlet parton densities on the lattice}
\vskip 1.6ex
\centerline
{\Bigbss in the Schr\"odinger Functional scheme}
\vskip 1.0 true cm
\centerline{\bigrm  Filippo Palombi$^{a}$, Roberto Petronzio$^a$, Andrea Shindler$^{a,b}$\kern1pt}
\vskip 0.8 true cm
\centerline{\it  $^a$Dipartimento di Fisica, Universit\`a di Roma Tor Vergata}
\vskip 0.15 true cm
\centerline{\it  and INFN, Sezione di RomaII,}
\vskip 0.3 true cm
\centerline{\it Via della Ricerca Scientifica 1, 00133 Rome, Italy} 
\vskip 0.6 true cm
\centerline{\it  $^b$Deutsches Elektronen--Syncrotron, DESY}
\vskip 0.3 true cm
\centerline{\it Platanenallee 6, D--15798 Zuthen, Germany} 
\vskip 1.0 true cm
\thintablerule
\vskip 2.0ex
\ninepoint
\leftline{\bf Abstract} 
\vskip 1.0ex\noindent
A non perturbative computation of the evolution of singlet parton densities without gauge--fixing requires 
a gauge invariant gluon source operator. Within the Schr\"odinger Functional scheme (SF), 
such a source can be defined in terms of path ordered products of gauge links, connected to 
the time boundaries. In this paper we adopt this definition and perform a 
one loop lattice computation of the renormalization constants of the twist--2 operators that
 correspond to the second moment of singlet parton densities. This calculation fixes the connection 
between the lattice SF scheme where a non perturbative evaluation of the absolute normalization of
singlet parton densities can be made at low energy and the $\overline{MS}$ scheme where one can extract 
the experimental values.
\vskip 0.3cm
Keywords: lattice QCD, perturbation theory, structure functions.

\vskip 2.0ex
\thintablerule
\vskip 8.0ex 
\centerline{March 2002}
\vfill\eject
\footline={\ifnum\pageno=0{\hfil}\else
{\hss\rm\the\pageno\hss}\fi}
\tenpoint
\section 1. Introduction

Non perturbative calculations of parton densities are the the only possibility to
fix their absolute normalization from first principles. Lattice simulations are
suitable to this purpose and various estimates have been presented in the literature
for the first moments of valence quark densities [\ref{strone},\ref{strtwo},\ref{strthree},\ref{strfour}]. 
In general, experimental values of moments
of the structure functions are  obtained by comparing production rates with theoretical cross
sections in common  continuum scheme at high energy, where perturbation  theory becomes reliable. 
For the non singlet case, the matching of low energy estimates of hadron matrix elements of Wilson
operators in lattice schemes with their experimental values has been realized within the
Schr\"odinger Functional scheme (SF), integrated by a finite size recursive method to match the large
gap of the energy scales involved. A crucial element of the calculation was the non perturbative
evolution of the Wilson operator taken in a matrix element with a proper SF quark {\it source}
that defined the SF scheme [\ref{gjp}]. An analogous calculation for the gluon density and in general for
the singlet parton densities involving the mixing between gluon and sea quark densities has not
yet been attempted, given the difficulties in accounting the sea quark pair creation
through unquenched simulation algorithms. 
In this paper we propose a defintion in the SF scheme of a gluon source that
can be used to evaluate the non perturbative running of the mixing renormalization matrix
characterising the singlet evolution. Using the Wilson action and the 
Feynman gauge, we perform a one loop calculation that fixes 
the relation between the SF singlet scheme and the $\overline{MS}$ scheme for the second moment of
singlet densities. 
Such a calculation is preliminary to the non perturbative evaluation of the
singlet densities hadron matrix elements and to their comparison with experimental
data. The method can be extended to higher moments where the experimental information is scarce [\ref{gluonex}]
and even a modeste precision can help fixing the gluon density at momentum fraction greater than
$0.5$ [\ref{procrob}].
\vskip 0.5cm
The paper is organized as follows: section 2 contains basic definitions of the SF scheme, singlet operators
and SF quark source. In section 3 a gauge invariant gluon source is introduced and its perturbative $O(g_0^2)$
expansion is performed. In section 4 we define correlation functions involving singlet operators, perform 
the one loop calculation and in section 5 we extract the one loop renormalization constants of the singlet operators.
\section 2. Singlet Structure Functions

\subsection 2.1 Schr\"odinger Functional

This section is only meant to recall some basic facts that have been discussed 
exhaustively in the literature [\ref{sffond}]. 
The theory is set up on a hyper-cubic euclidean lattice with spacing $a$ and size
$T\times L^3$ (throughout the paper we put $T=L$, writing $T$ whenever it has to be
recalled the time character of the variable).
The Schr\"odinger functional represents the amplitude for the time evolution 
that takes into account quantum fluctuations of a classical
field configuration between two predetermined classical states.
It takes the form of a standard functional integral with fixed boundary 
conditions.
Explicitly, the link variables are chosen to be periodic in space and to satisfy Dirichlet
boundary conditions in time,
\equation{
A_k(x)|_{x_0=0}=C(\bx),\qquad
A_k(x)|_{x_0=T}=[\projector C'](\bx)\,,
\enum}
where $C(\bx)$ and $C'(\bx)$ are fixed boundary fields,
and $\projector$ projects onto the gauge invariant content of $C'(\bx)$
[\ref{sffond}]. Here we choose $C=C'=0$,
leading to the boundary conditions
\equation{
  U(x,k)|_{x_0=0}=\su3id\,,\quad
  U(x,k)|_{x_0=T}=\su3id\,;\qquad k = 1,2,3
\enum}
for the lattice gauge field. The only gauge transformations on the boundary time slices
that preserve such boundary conditions are the global ones [\ref{sffond}]. This property will be
 crucial for the definition of the gauge invariant surface sources.
The quark fields are chosen to be periodic up to a phase in the
three space directions,
\equation{
  \psi(x+L\hk)=\e^{i\theta_k}\psi(x),
  \qquad
  \psibar(x+L\hk)=\psibar(x)\e^{-i\theta_k}\,;\qquad k = 1,2,3
\enum}
where $\theta_k$ is kept as a free parameter.
One can also distribute the phase to all lattice points 
by an abelian transformation on the Fermi fields,
i.e. by changing the form of the usual lattice derivative in
\equation{
  \nabla_\mu\psi(x)={1\over a}[\lambda_\mu
  U(x,\mu)\psi(x+a\hmu)-\psi(x)]
\enum}
\equation{
  \nabla_\mu^*\psi(x)={1\over a}[
  \psi(x) - \lambda_\mu^{-1}U(x-a\hmu,\mu)^{-1}\psi(x-a\hmu)]
\enum}
\noindent where
\equation{
  \lambda_\mu = \e^{ia\theta_\mu / L}\,; \qquad \qquad \theta_0 = 0\,, \quad
   -\pi < \theta_k \leq \pi
\enum}

It is useful to define also the backward derivatives on the lattice
\equation{
  \psibar(x) \overleftarrow \nabla_\mu = 
  {1\over a}[\psibar(x + a \hmu)U(x,\mu)^{-1}\lambda_\mu^{-1} - \psibar(x)]
\enum}
\equation{
   \psibar(x) \overleftarrow \nabla_\mu^{*} = 
   {1\over a}[\psibar(x) - \psibar(x - a \hmu)U(x - a\hmu,\mu)\lambda_\mu]
\enum}
Similarly to the gauge field, Dirichlet boundary conditions are imposed on
the quark fields,
\equation{
  P_{+}\psi(x)|_{x_0=0}=\rho(\bx),\quad
  P_{-}\psi(x)|_{x_0=T}=\rho'(\bx),
\enum}
and
\equation{
  \psibar(x)P_{-}|_{x_0=0}=\rhobar(\bx),\quad
  \psibar(x)P_{+}|_{x_0=T}=\rhobar'(\bx),
\enum}
The Schr\"odinger functional action, including Feynman rules for the fermionic
part and further details can be found in~[\ref{sffond},\ref{pert}].
\subsection 2.2 Singlet operators

In the continuum, moments of singlet structure functions are related, 
through the operator product expansion, to hadronic matrix elements of two kind of   
twist--2, gauge invariant, local operators of the form

\equation{\eqalign{
  & O^q_{\mu_1 \cdots \mu_N} = {1\over 2^N}\bar\psi \gamma_{[\mu_1} 
  \lrD_{\mu_2} \cdots \lrD_{\mu_N]}
  \psi \cr
  & O^g_{\mu_1 \cdots \mu_N} = \sum_{\rho}\tr\bigl\{F_{[\mu\rho}
  \lrD_{\mu_2} \cdots \lrD_{\mu_{N-1}}
  F_{\rho\mu_N]}\bigr\}
  }
\enum}
where brackets [ \dots ] mean Lorentz indices symmetrization
and $\lrD_\mu = \rD_\mu - \lD_\mu$, with
\equation{
  \rD_\mu = {1\over 2}(\nabla_\mu + \nabla^*_\mu), 
  \qquad 
  \lD_\mu = {1\over 2}
  (\overleftarrow\nabla_\mu + \overleftarrow\nabla^*_\mu)
\enum}
The operators in eq.~(2.11) belong  in the continuum to 
irreducible representations of the angular momentum.
On the lattice, given the lower (hypercubic) 
symmetry of the Euclidean lattice action with respect 
to that of the continuum (all 4--d rotations), 
the identification of an irreducible representation may require some 
particular combination of operators. 
This classification has been discussed for
example in refs. [\ref{grone},\ref{grtwo}]. 
A subset of the basis described in [\ref{grtwo}], involving
only spatial indices is given by
\equation{
  O^q_{12}(x) = {1\over 4}\psibar(x) \gamma_{[1} 
  \lrD_{2]}\psi(x)
\enum}
\equation{
  O^g_{12} = \sum_{\rho} \tr\{F_{[1\rho}(x) F_{\rho 2]}(x)\} 
\enum}
and 
\equation{
  F_{\mu\nu}(x)={1\over 8 a^2 g_0}\left\{
  Q_{\mu\nu}(x)-Q_{\nu\mu}(x)\right\}
\enum}
\midinsert
\vbox{
\vskip0.0cm
\epsfxsize=5.0cm\hskip3.5cm\epsfbox{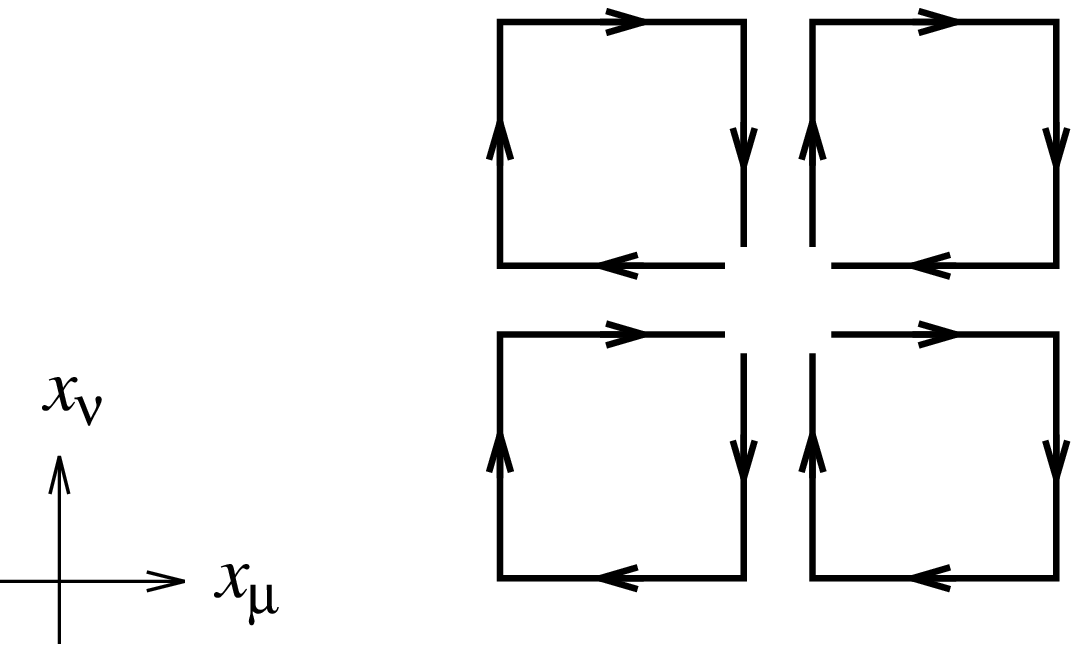}
\vskip0.2cm
\figurecaption{%
\eightrm Graphical representation of the products of gauge field
variables contributing to the lattice field tensor, eq.~(2.15). The point $x$
is at the center of the diagram where all loops start and end.
}
\vskip0.0cm
}
\endinsert
with $Q_{\mu\nu}(x)$ being the sum of the plaquette loops shown in Fig.~1, 
maximizing the symmetry of the field strength on the lattice. The usual 
plaquette representation of the field strength transforms like a reducible representation of the
hypercubic group, while the definition {\it clover}--like of eq.~(2.15)
tranfsorms like an irreducible one.

The correlation functions we want to study can be organized in a $2\times 2$ matrix as follows 
\equation{
  f_{\alpha\beta} \sim \left[\matrix{<{\cO}_q~ \Source_q> & <{\cO}_q~ \Source_g> \cr %
				     <{\cO}_g~ \Source_q> & <{\cO}_g~ \Source_g> \cr} \right]
\enum}
where $\cO$'s are the operators and $\Source$'s 
are the quark and gluon sources, defined in the following sections.
Beyond the tree--level there will be in general a mixing in the flavor singlet
sector between the quark operator (2.13) and the gluon operator (2.14).
However in the {\it{quenched}} approximation ($N_f = 0$) one has $<{\cal O}^q_{\mu \nu}~\Source_g >=0$.
In this case the matrix (2.16) becomes triangular and the operator ${\cal O}^q$ does not mix
with ${\cal O}^g$.
\subsection 2.3 Fermionic source

To probe operators (2.13) and (2.14) in correlation functions,
one must choose suitable sources.
In the case of the quark source we make the same choice of the
non singlet calculation [\ref{bucpal}].
As already stressed in that paper, the SF scheme allows us to define 
a gauge invariant source that provides a spatial direction.
The operator (2.13) needs two directions and we have to give to the quark state 
a momentum different from zero in one extra direction.
Moreover, using a particular feature of the SF, we can use the constant phase term
 $\btheta$ defined in (2.3), called a {\it finite--size} momentum.
The reason for this name is that, at finite volume, this phase acts like a momentum probed 
by the local operators that we want to renormalize,
but unlike the standard lattice momentum, it escapes the quantization rule induced by the
 finite volume. Its value can be chosen smaller than the minimum value of
the standard momentum $p_{min} = {2\pi \over L}$, reducing the associated important
lattice artefacts [\ref{bucpal}]. The quark bilinear at the 
boundary is given by the expression
\equation{
  \Source_q(\bp) = a^6 \sum_{\by,\bz}\e^{i\bp\cdot(\by - \bz)}\ \bar\zeta(\by)\gamma_2\zeta(\bz)
\enum}

\section 3. Gauge invariant gluon source

In this section we introduce a gauge invariant gluon source which 
can be used for the calculation of the singlet correlation functions [\ref{shitalk}].
To exploit the features of the SF we recall that the gauge group of the
 SF is local in the bulk and global on the boundaries.
This makes the quark source (2.17) gauge invariant. The gauge invariant gluon source
is defined by
\equation{
  \Source_g = {\cal S} = \tr\{{\cal T}_1 {\cal T}_2 \}
\enum}
where the trace is over the color indices. The {\it big--tooth} state ${\cal T}_i$ is defined by
\equation{
  {\cal T}_i = {a^3\over 2i}\sum_\bx\biggl\{\Pi_i(\bx)-\Pi_i^{\dagger}(\bx)\biggr\}
\enum}
and
\equation{
  \Pi_i(\bx) = {1\over ag_0}\prod_{x_0=0}^{{T\over 4}-a}\ U_0(x_0,\bx) \ U_i\biggl({T\over 4},\bx\biggr)
  \prod_{x_0={T\over 4}-a}^0U^{-1}_0(x_0,\bx+a\hi)
\enum}
It is natural to define a gluon source also associated with the boundary $x_0 = T$.
\equation{
  {\cal S}' = \tr\{{\cal T}_1'{\cal T}_2'\}
\enum}
\equation{
  {\cal T}_i' = 
  {a^3\over 2i}\sum_\bx\biggl\{\Pi_i'(\bx)- {\Pi'}_i^{\dagger}(\bx)\biggr\}
\enum}
\equation{
  \Pi_i'(\bx) = 
  {1\over ag_0}\prod_{x_0={T-a}}^{{3\over 4}T}\ U_0(x_0,\bx) \ 
   U_i\biggl({{3\over 4}T},\bx\biggr)
  \prod_{x_0={{3\over 4}T}}^{T-a}U^{-1}_0(x_0,\bx+a\hi)
\enum}
Eq.~(3.4) is necessary in order to define the correlation function which will 
be used to remove the additional divergences of the singlet correlation
functions, introduced by the source. 
A graphical representation of the source is reported in Fig.~2.
It is a product of temporal 
links in the time direction from $x_0 = 0$ to $x_0=T/4$, 
connected with a spatial link, or from $x_0 = T$ to $x_0 = 3T/4$.
It is gauge invariant, has two spatial directions (the same ones of the operators
 (2.13) and (2.14)), and is projected at zero momentum. 
The gluon source at $x_0 = T$ cannot be 
obtained by substituting in ${\cal S}$ $x_0=T/4$ with $x_0 = 3T/4$.
The source gives raise to linear and logarithmic divergences: our calculation in perturbation theory shows
that they are removed by a proper normalization of the correlation
functions where it appears.
\midinsert
\vbox{
\vskip0.0cm
\epsfxsize=3.0cm\hskip6.0cm\epsfbox{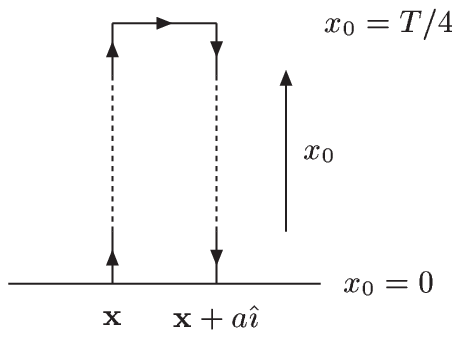}
\vskip0.2cm
\hskip 4.5cm\figurecaption{%
\eightrm Graphical representation of $\Pi_i(\bx )$.
}
\vskip0.0cm
}
\endinsert

\subsection 3.1 Perturbative expansion of the gluon source

The perturbative expansion of the gluon source is straightforward.
In the SF we adopt as usual the time--momentum scheme defined in [\ref{pert}].
For the color matrices we follow the same convention of [\ref{pert}].
So we have
\equation{
  U(x,\mu) = \exp\bigl\{ag_0q_{\mu}^a(x)T^a\bigr\}
\enum}
\equation{
  q_0^a(x) = {1\over L^3} \sum_{\bp} {\rm{e}}^{i\bf{p}\cdot \bf{x}} 
  \tilde q_0^a(x_0;{\bf{p}}) 
\enum}
\equation{
  q_k^a(x) = {1\over L^3} \sum_{\bf{p}} {\rm{e}}^{i\bf{p}\cdot  \bf{x}} 
  {\rm{e}}^{{i\over 2}ap_k}\tilde q_k^a(x_0;{\bf{p}}) 
\enum}
The sum over the momenta {\bp} runs in the range $-\pi/a < p_k \leq \pi/a$ with
\equation{
  {\bp} = (p_1,p_2,p_3), \quad p_k = {2\pi\over L}n_k,\quad n_k\in \Zset  
\enum}
The Feynman rules associated with the gluon sources are found expanding in powers of $g_0$
\equation{
  {\cal T}_i \ = \ {\cal T}_i^{(0)} + g_0{\cal T}_i^{(1)} + g_0^2{\cal T}_i^{(2)}
\enum}
\equation{
  {\cal T}_i' \ = \ {\cal T}_i'^{(0)} + g_0{\cal T}_i'^{(1)} + g_0^2{\cal T}_i'^{(2)}
\enum}
Each term of the expansion is given by the sum of various contributions 
that we enumerate alphabetically with big case latin letters (where needed). Here are the tree--level expressions:
\equation{
  {\cal T}_i^{(0)} \ = \ -i\tilde q_i^a\biggl({T\over 4}, \bzero \biggr)T^a
\enum}
\equation{
  {\cal T}_i'^{(0)} \ = \ -i\tilde q_i^a\biggl({3T\over 4}, \bzero \biggr)T^a
\enum}
where repeated indices are summed. The $O(g_0)$ terms are:
\equation{
  {\cal T}_{i}^{(1,A)}  = {i\over L^3}\sum_\bp 
  a \sum_{u_0=0}^{T/4-a} \cos\biggl({a\over 2}p_i\biggr)\ \tilde q_i^a\biggl({T\over 4},\bp\biggr)\
  \tilde q_0^b(u_0,-\bp)\ f^{abc}T^c
\enum}
\equation{
  {\cal T}_{i}^{(1,B)} = {1\over 2L^3}\sum_\bp a^2 \sum_{u_0,v_0=0}^{T/4-a} \ppal_i\ 
  \tilde q_0^a(u_0,\bp)\ \tilde q_0^b(v_0,-\bp)\ f^{abc}T^c
\enum}
\equation{
  {\cal T}_{i}'^{(1,A)} = 
  - {i\over L^3}\sum_\bp  
  a \sum_{u_0=\ 3T/4}^{T-a}\cos\biggl({a\over 2}p_i\biggr)\
  \tilde q_i^a\biggl({3T\over 4},\bp\biggr)\ 
  \tilde q_0^b(u_0,-\bp)\ f^{abc}T^c\
\enum}
\equation{
  {\cal T}_{i}'^{(1,B)} = 
  {1\over 2L^3}\sum_\bp a^2 \sum_{u_0,v_0=\ 3T/4}^{T-a} \ppal_i\ 
  \tilde q_0^a(u_0,\bp)\ \tilde q_0^b(v_0,-\bp)\ f^{abc}T^c
\enum}
And the $O(g_0^2)$ are given by:
\equation{
  {\cal T}_{i}^{(2,A)} = -{i\over 6}{a^2\over L^6}\sum_{\bp,\bq}\tilde q_i^a\biggl({T\over 4},\bp\biggr)\ 
  \tilde q_i^b\biggl({T\over 4},\bq\biggr)\ \tilde q_i^c\biggl({T\over 4},-\bp-\bq\biggr) T^aT^bT^c
\enum}
\equation{
  {\cal T}_{i}^{(2,B)} =
  {i\over L^6}\sum_{\bp,\bq} a^2 \sum_{u_0,v_0=0}^{T/4-a} 
  \cos\biggl[{a\over 2}(p_i-q_i)\biggr]\ \tilde q_0^a(u_0,\bp)\tilde q_i^b\biggl({T\over 4},-\bp-\bq\biggr) 
  \tilde q_0^c(v_0,\bq)\ T^aT^bT^c
\enum}
\equation{\eqalign{
  {\cal T}_{i}^{(2,C)} & = - {i\over L^6}\sum_{\bp,\bq} a^2 \sum_{u_0\le v_0=0}^{T/4-a}
  c(u_0,v_0)\ \biggl\{ \cos\biggl[{a\over 2}(p_i+q_i)\biggr]\
  \tilde q_0^a(u_0,\bp) \tilde q_0^b(v_0,\bq)\ \times \cr
  & \times \ \tilde q_i^c\biggl({T\over 4},-\bp-\bq\biggr) +
  \cos\biggl({a\over 2}p_i\biggr)\ \tilde q_i^a\biggl({T\over 4},\bp \biggr)
  \tilde q_0^b(v_0,\bq)\ \tilde q_0^c(u_0,-\bp -\bq) \biggr\} T^aT^bT^c
  }
 \enum}
\equation{
  {\cal T}_{i}'^{(2,A)}  = 
  -{i\over 6}{a^2\over L^6}
  \sum_{\bp,\bq}\tilde q_i^a\biggl({3T\over 4},\bp\biggr)\ 
  \tilde q_i^b\biggl({3T\over 4},\bq\biggr)
  \tilde q_i^c\biggl({3T\over 4},-\bp-\bq\biggr)\ T^aT^bT^c
\enum}
\equation{\eqalign{
  {\cal T}_{i}'^{(2,B)} & =
  {i\over L^6}\sum_{\bp,\bq} a^2 \sum_{u_0,v_0=\ 3T/4}^{T-a} \cos\biggl[{a\over 2}(p_i-q_i)\biggr]
  \tilde q_0^a(u_0,\bp)\tilde q_i^b\biggl({3T\over 4},-\bp-\bq\biggr) \times \cr
  & \times \ \tilde q_0^c(v_0,\bq)\ T^aT^bT^c}
\enum}
\equation{\eqalign{
  {\cal T}_{i}'^{(2,C)} & = - {i\over L^6}\sum_{\bp,\bq} a^2 \sum_{u_0\ge v_0=\ 3T/4}^{T-a}c(u_0,v_0)\ \biggl\{ 
  \cos\biggl[{a\over 2}(p_i+q_i)\biggr]
  \tilde q_0^a(u_0,\bp) \tilde q_0^b(v_0,\bq)\ \times \cr 
  & \times \ \tilde q_i^c\biggl({3T\over 4},-\bp-\bq\biggr) + 
  \cos\biggl({a\over 2}p_i\biggr) \tilde q_i^a\biggl({3T\over 4},\bp\biggr)
  \tilde q_0^b(v_0,\bq)\ \tilde q_0^c(u_0,-\bp-\bq)\biggr\} T^aT^bT^c
  }
\enum}
where 
\equation{
c(u_0,v_0) = \left\{\matrix{1 & {\rm{if}} \quad u_0 \neq v_0\cr
			   {1\over 2} & {\rm{if}} \quad u_0 = v_0}\right.
\enum}
\midinsert
\vbox{
\vskip0.0cm
\epsfxsize=11.0cm\hskip1.7cm\epsfbox{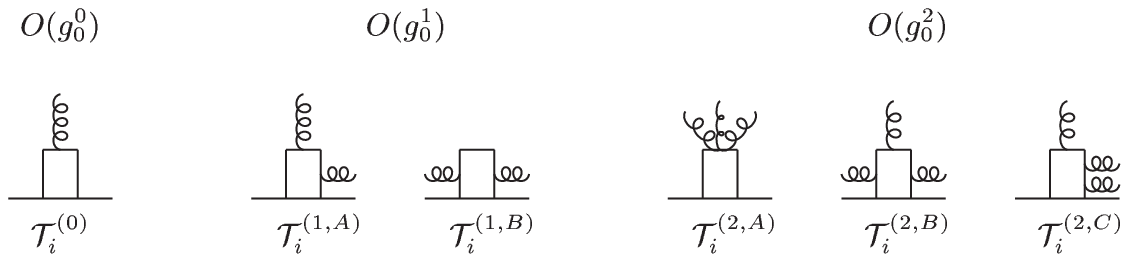}
\vskip0.2cm
\hskip 2.2cm\figurecaption{%
\eightrm The contributions to the perturbative expansion of ${\cal T}_i$ up to $O(g_0^2)$.
}
\vskip0.0cm
}
\endinsert
A graphical representation of the Feynman rules for the source
at $x_0 = 0$ is reported in Fig.~3 and the source at $x_0=T$ has an analogous graphical interpretation.
From (3.13) and (3.14) one can see that the tree--level of the gluon sources is a gluon field
at time $x_0 = T/4$ or $x_0 = 3T/4$ with spatial polarization and zero momentum.
We have now all the ingredients that allow to perform a perturbative one loop calculation of
the renormalization constants of the singlet operators and of the correlation functions involving
only the sources.

\section 4. Renormalization

The renormalization condition connects the bare operators on the lattice 
to finite operators renormalized at a scale $\mu = 1/L$ :
\equation{
  {\cal O}_l^R(\mu) = Z_{lk}(\mu a){\cal O}_k(a)
\enum}
As we have discussed above, in the flavor singlet sector (and in the $unquenched$ case) 
there is a mixing between the quark (2.13) and the gluon operator (2.14) with the same quantum numbers.
We thus write
\equation{
  {\cal O}_q^R = Z_{qq}{\cal O}_q + Z_{qg}{\cal O}_g
\enum}
\equation{
  {\cal O}_g^R = Z_{gq}{\cal O}_q + Z_{gg}{\cal O}_g
\enum}
We adopt the following renormalization conditions
in the SF scheme:
\equation{
  <{\cal O}_q^R(\mu)~\Source_q>|_{\mu = 1/L}~=~<{\cal O}_q(a)~\Source_q>|^{tree}
\enum}
\equation{
  <{\cal O}_q^R(\mu)~\Source_g>|_{\mu = 1/L}~=~<{\cal O}_q(a)~\Source_g>|^{tree}~=~0
\enum}
\equation{
  <{\cal O}_g^R(\mu)~\Source_q>|_{\mu = 1/L}~=~<{\cal O}_g(a)~\Source_q>|^{tree}~=~0
\enum}
\equation{
  <{\cal O}_g^R(\mu)~\Source_g>|_{\mu = 1/L}~=~<{\cal O}_g(a)~\Source_g>|^{tree}
\enum}
From eqs.~(4.2) -- (4.3) and (4.4) -- (4.7), the
renormalization constants can be written for $a/L \ll 1$ in the form 
[\ref{gross}]
\equation{
  Z_{qq}(a/L)~=~1 - {\alpha_S\over 4\pi}C_F\Biggl[{16\over 3} \log{a\over L} + B_{qq} + O\biggl({a\over L}\biggr)\biggr]
 +  O(\alpha_S^2)
\enum}
\equation{
  Z_{qg}(a/L)~=~-{\alpha_S\over 4\pi}N_f\biggl[{4\over 3} \log{a\over L} + B_{qg} +O\biggl({a\over L}\biggr)\biggr]
+ O(\alpha_S^2)
\enum}
\equation{
  Z_{gq}(a/L)~=~-{\alpha_S\over 4\pi}C_F\biggl[{16\over 3} \log{a\over L} + B_{gq} + O\biggl({a\over L}\biggr)\biggr]
 + O(\alpha_S^2)
\enum}
\equation{
  Z_{gg}(a/L)~=~1 - {\alpha_S\over 4\pi}\Big[N_f\Big({4\over 3} \log{a\over L} + B^f_{gg}\Big)
  + N_c B^g_{gg} + O\biggl({a\over L}\biggr)\Big] + O(\alpha_S^2)
\enum}
where $C_F = 4/3$ and $\alpha_S = g^2/{4\pi}$. The coefficients of the logarithms represent
the  {\it anomalous dimensions} of the corresponding operators. They are responsible for the
RG--evolution of the coefficient functions. The $B$'s are fixed by the renormalization conditions
of eqs.~(4.4) -- (4.7).

\subsection 4.1 Correlation functions

In order to define the correlation functions for the calculation 
of the singlet renormalization constants, according to eq.~(2.16), we introduce the following
four correlation functions
\equation{
  f_{qq}(x_0,\bppiu) = -a^6\sum_{\bf{y},\bf{z}} 
  \rm{e}^{i{\bf{p}}\cdot ({\bf{y}}-{\bf{z}})} 
  \bigl\langle {1\over 4} \bar\psi(x) \gamma_{[1} 
  \lrD_{2]}\psi(x) \bar\zeta({\bf{y}}) \gamma_2 \zeta({\bf{z}})\bigr\rangle
\enum}
\equation{
  f_{qg}(x_0,\bppiu) = -a^6\sum_{\bf{y},\bf{z}} 
  \rm{e}^{i{\bf{p}}\cdot ({\bf{y}}-{\bf{z}})}  
  \bigl\langle \sum_{\rho}\tr\{F_{[1\rho}(x) F_{\rho2]}(x)\} 
  \bar\zeta({\bf{y}}) \gamma_2 \zeta({\bf{z}})\bigr\rangle
\enum}
\equation{
  f_{gq}(x_0) =  
  \bigl\langle {1\over 4} \bar\psi(x) \gamma_{[1} 
  \lrD_{2]}\psi(x) \tr\{{\cal T}_1 {\cal T}_2\}\bigr\rangle
\enum}
\equation{
  f_{gg}(x_0) = \bigl\langle  \sum_{\rho}\tr\{F_{[1\rho}(x) F_{\rho2]}(x)\} 
  \tr\{{\cal T}_1 {\cal T}_2\}\bigr\rangle
\enum}
In the correlation functions (4.12) and (4.13) we perform the computation with ${\bf{p}} = 0$ and
$\btheta = (\theta_1,0,0)$. Moreover, we are free to choose 
the physical distance $x_0$ of the operator insertions from the lower boundary, and we fix $x_0 = T/2$ in all
the correlation functions. We have also to define correlation functions 
which involve the sources both at $x_0 = 0$ and $x_0 = T$,
because the correlation functions (4.12) -- (4.15) have to be properly 
normalized by removing the renormalization of the sources.
The quark source has a well known logarithmic divergence [\ref{pert},\ref{bucpal}].
Our gluon source, as it will be seen, has a leading linear divergence. 
The correlation function for the quark source is
\equation{
  f_1 = - {a^{12}\over L^6}\sum_{\bu,\bv,\by,\bz}\langle{\bar{\zeta}}'(\bu)\gamma_5 \zeta'(\bv)
  {\bar{\zeta}}(\by) \gamma_5 \zeta(\bz)\rangle
\enum}
and for the gluon source:
\equation{
  G_1 = {1\over L^8}\langle {\cal S} {\cal S}' \rangle
\enum}
where ${\cal S}$ and ${\cal S}'$ are defined in 
(3.1) and (3.4).
We can calculate analytically the tree--level of the diagonal correlation
functions $f_{qq}$ and $f_{gg}$ and of the sources correlation
functions $f_1$ and $G_1$.
The tree--level of $f_1$ with $\bp = \btheta = 0$ and $m_0 = 0$ is
\equation{
  f_1^{(0)} = N_c
\enum}
where $N_c$ is the number of colors.
The tree--level of $G_1$ is
\equation{
  G_1^{(0)} \ = \ \biggl({T\over L}\biggr)^2\ {N_c^2 - 1\over 1024}
\enum}
The tree--level of $G_1$ is a constant and therefore the same on
the lattice and on the continuum. 
The correlation functions with a non zero tree--level are $f_{qq}(x_0;\theta_1)$ 
and $f_{gg}(x_0)$. The {\it tree--level} of $f_{qq}$ is
\equation{
  f^{(0)}_{qq}(x_0;\theta_1) = {i\ppal_1^+ N_c\over R(p^+)^2} \left[ (-i\ppal_0) 
  \left(M_-(p^+) {\rm{e}}^{-2\omega({\bf{p}}^+)x_0} - 
  M_+(p^+) {\rm{e}}^{-2\omega({\bf{p}}^+)(2T-x_0)}\right) \right ]\biggr|_{\bp = 0}
\enum}
and in the continuum and chiral limit it takes the form
\equation{
  f^{(0)}_{qq}(x_0;\theta_1) = {\theta_1\over L}{N_c \over (1+\e^{-2 {\theta_1\over L} T})^2}
  \left[\e^{-2 {\theta_1\over L} x_0} + \e^{-2 {\theta_1\over L} (2T-x_0)} \right].
\enum}
The purely gluonic correlation function has a tree--level of the form
\equation{\eqalign{
  f^{(0)}_{gg}(x_0) = {\ N_c^2 - 1\ \over 16a^2} &
  [d_{11}(x_0 + a, T/4;\bzero) - d_{11}(x_0 - a, T/4;\bzero)] \times  \cr
  \times \ & [d_{22}(x_0 + a, T/4;\bzero) - d_{22}(x_0 - a, T/4;\bzero)] 
}
\enum}
where $d_{\mu\mu}( y_0, z_0; \bq )$ is the time--momentum gluon propagator connecting $y_0$ and $z_0$ time slices, carrying
a momentum $\bq$ and polarization $\mu$ [\ref{pert}]. Eq. (4.22) is the square of the time lattice derivative of the spatial gluon propagator. The fact that the spatial gluon propagator with zero momentum
is linear in time coordinates [\ref{pert}], implies that with 
$x_0 \pm a > T/4$, the expression of $f^{(0)}_{gg}(x_0)$ is
independent of $x_0$  and reads
\equation{
  f^{(0)}_{gg}(x_0) \ = \ {\ N_c^2 - 1 \ \over 64}
\enum}
Lattice and continuum expressions are identical. 

\subsection 4.2 One loop perturbative expansion

The only missing ingredients to perform a one loop perturbative expansion 
of the correlation functions (4.12) -- (4.15) and (4.16) -- (4.17)
are the Feynman rules coming from the expansion of the
operators and from the gauge part of the action.
In Appendix A we give the perturbative expansion of the field strength $F_{\mu \nu}$ to 
first order. The second order is rather cumbersome and not so instructive.
In Appendix B we give the Feynman rules of the gauge part of the action.

The expansion of $f_1$ was already done in [\ref{pert}], while
the expansion of $f_{qq}$ is actually the same performed in the non singlet calculation [\ref{bucpal}].
In the following, we will restrict the attention to the new correlation functions.
\vskip 0.3cm
\hskip 0.5cm {\it i)} \ $G_1$

The correlation function involving only gluon sources, $G_1$, defined in (4.17), can be expanded in terms of
the perturbative expansion of ${\cal S}$, ${\cal S}'$ and the action. The Feynman diagrams of the expansion of
this correlation function are given in Fig.~4. There are two checks of the calculation.
The gluon self--energy develops quadratic
divergences that cancel out by summing all the self--energy contributions
(cfr. diagrams 6.a -- 6.g in Fig.~4), and the cancellation has to take place separately in the gluon
sector (cfr. diagrams 6.a -- 6.e) and in the fermion one (cfr. diagrams 6.f -- 6.g).
This is verified in our computation.
\topinsert
\vbox{
\vskip1.0cm
\epsfxsize=9cm\hskip3.2cm\epsfbox{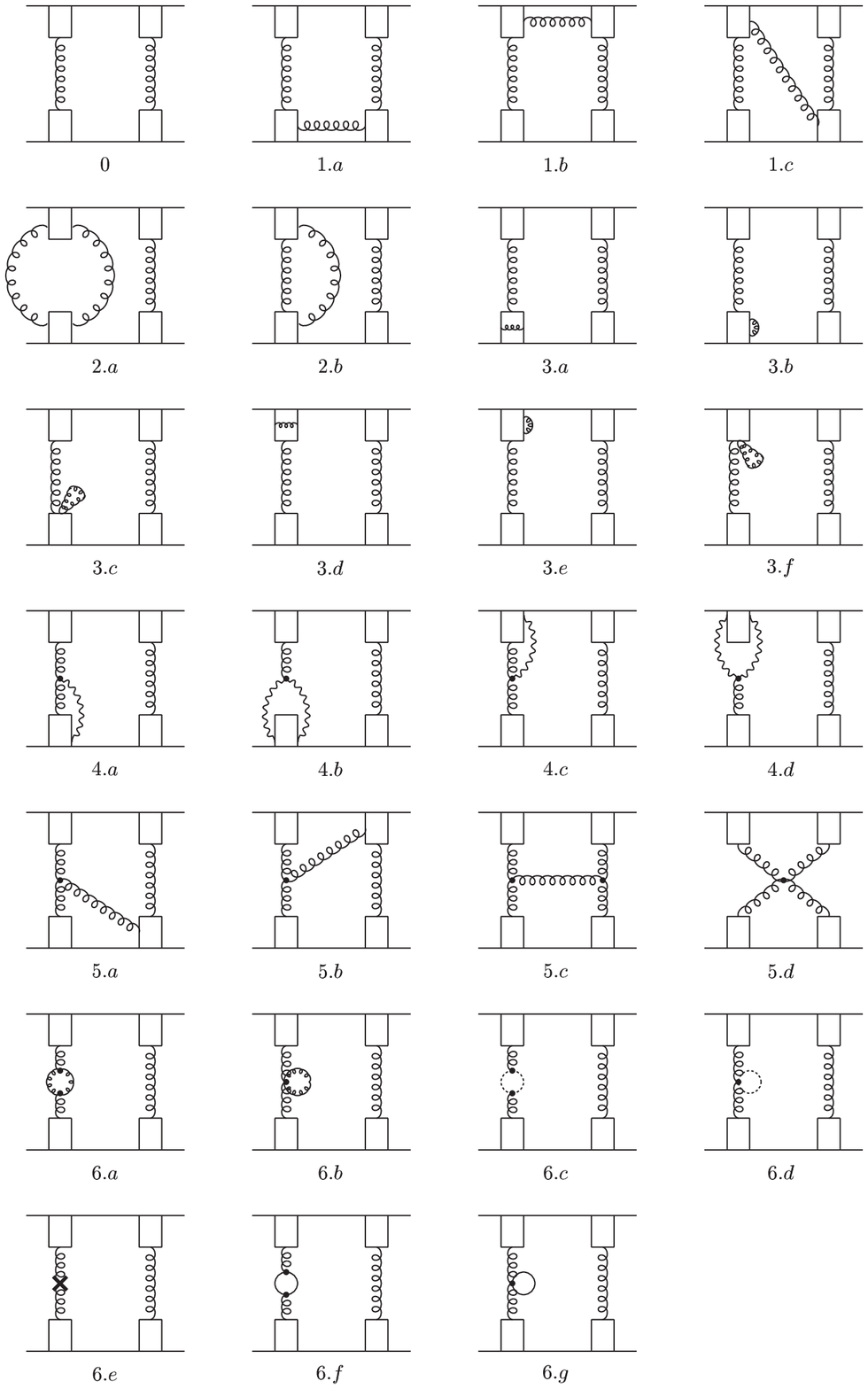}
\vskip0.2cm
\hskip 3.5cm\figurecaption{%
Feynman diagrams of the one loop expansion of $G_1$.
}
\vskip 0.0cm
}
\endinsert
The second check is specific to our definition of the  gluon source.
As anticipated in section 2, the source develops linear divergences
that must cancel once we normalize $f_{gg}$ with $\sqrt{G_1}$ and they do so.  Numerical results for the ratio
\equation{
  {G_1^{(1)}\over G_1^{(0)}} = R_1^{(gl)} + N_f R_1^{(fe)}
\enum}
are reported in Table~1, wher it can be seen the presence
of linear divergences in $R_1^{(gl)}$.
\vskip 0.3cm
\hskip 0.5cm {\it ii)} \ $f_{qg}$

The correlation function $f_{qg}$, concerning the gluon operator with the quark source, eq.~(4.13), can be expressed
through [\ref{pert}]
\equation{\eqalign{
[\zeta({\bf{x}})\bar{\zeta}({\bf{y}})]_F & = P_- U_0(x-a\hat{0})S(x,y)
U_0(y-a\hat{0})P_+|_{x_0=y_0=a} + \cr 
& -{1\over 2} P_- \gamma_k(\partial_k + 
\partial^*_k)a^{-2} \delta_{{\bf{x}}{\bf{y}}}
}
\enum}
which has to be expanded to the $O(g_0^2)$ order.
It also requires the tree--level of the gluon operator in the
time--momentum scheme, which can be easily computed through the
field strength expression given in Appendix A.
A picture of the Feynman diagrams for $f_{qg}$ is reported in Fig.~5.
For the computation of this correlation function, as in the case
of $f_{qq}$ [\ref{bucpal}], we use a {\it finite--size} momentum $\btheta = (0.1,0,0)$. 
\midinsert
\vbox{
\vskip2.0cm
\epsfxsize=6.0cm\hskip4.8cm\epsfbox{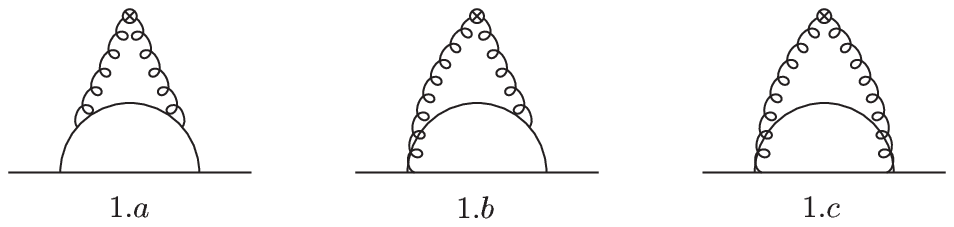}
\vskip0.2cm
\hskip 3.5cm\figurecaption{%
Feynman diagrams of the one loop expansion of $f_{qg}$.
}
\vskip 0.0cm
}
\endinsert
The expressions of the Feynman diagrams are quite involved, not so 
instructive and will not be shown.
Numerical results of $f^{(1)}_{qg}/f^{(0)}_{qq}$ are reported in Table~2.
\vskip 0.3cm
\hskip 0.5cm {\it iii)} \ $f_{gq}$

The second non diagonal correlation function, $f_{gq}$ is defined through eq.~(4.14).
Its computation involves only the tree--level of the gluon source.
It depends on $x_0$ but not on the momentum, because the gluon source is projected
to zero momentum. The expansion of the quark operator was already done in [\ref{bucpal}].
Feynman diagrams are depicted in Fig.~6. Numerical results are parametrized as
\midinsert
\vbox{
\vskip2.0cm
\epsfxsize=3.0cm\hskip5.8cm\epsfbox{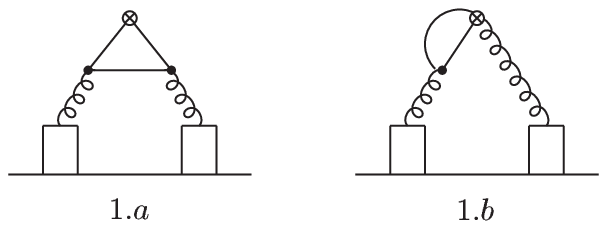}
\vskip0.2cm
\hskip 3.1cm\figurecaption{%
Feynman diagrams of the one loop expansion of $f_{gq}$.
}
\vskip 0.0cm
}
\endinsert
\equation{
{f_{gq}^{(1)}\over f_{gg}^{(0)}} = N_f R_{gq}^{(fe)}
\enum}

because this correlation function is proportional to the number of dynamical fermions.
Note that the value of $f^{(1)}_{qg}/f^{(0)}_{qq}$ is available for lattices with size multiple of two
while $R_{gq}^{(fe)}$ is available for lattices with size multiple of four, because
correlation functions with the gluon source involve $T/4$ as an integer parameter, while for the 
quark source they do not. Numerical results are reported in Table~2.
\vskip 0.3cm
\hskip 0.5cm {\it iv)} \ $f_{gg}$
The calculation of this gluonic correlation function, eq.~(4.15), is technically 
difficult. We have to use the expansion of the gluon source, eqs.~(3.13) -- (3.21),
and of the operator up to the second order.
We can do here the same checks that we have done for $G_1$.
Indeed, the quadratic divergences cancel out by summing all the self--energy contributions
(cfr. diagrams 5.a -- 5.g in Fig.~7), and the cancellation takes place separately in the gluon 
sector (cfr. diagrams 5.a -- 5.e) and in the fermion one (cfr. diagrams 5.f -- 5.g).
Linear divergences cancel out once we normalize $f_{gg}$ with $\sqrt{G_1}$.
Numerical results are presented in Table~3 by parametrizing them as follows
\equation{
{f_{gg}^{(1)}\over f_{gg}^{(0)}} = R_{gg}^{(gl)} + N_f R_{gg}^{(fe)}
\enum}
\topinsert
\vbox{
\vskip1.0cm
\epsfxsize=9.0cm\hskip3.0cm\epsfbox{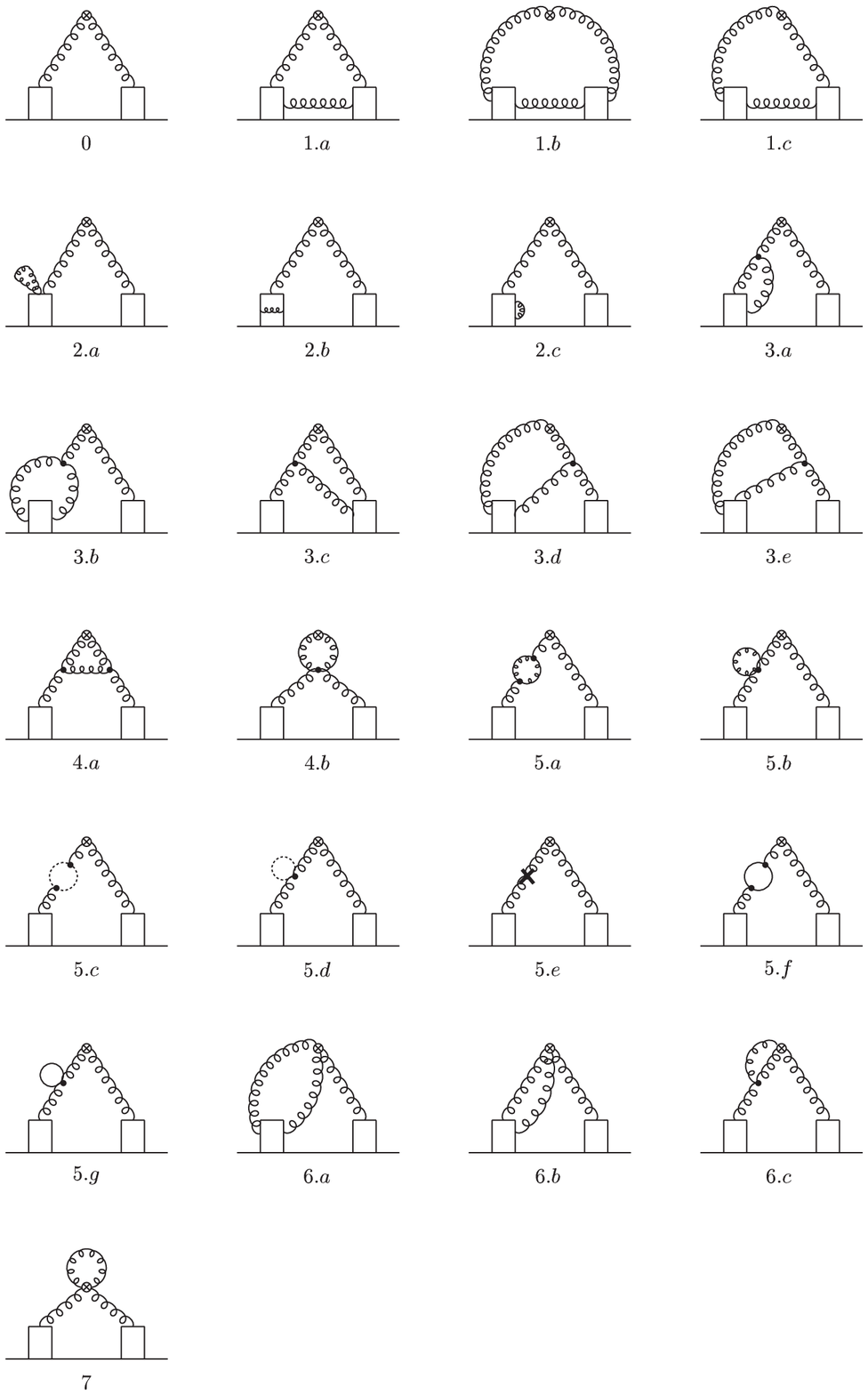}
\vskip0.2cm
\hskip 3.3cm\figurecaption{%
Feynman diagrams of the one loop expansion of $f_{gg}$.
}
\vskip 0.0cm
}
\endinsert
Note that $R_{gg}^{(gl)}$ has a linear divergence while $R_{gg}^{(fe)}$ has only
a logarithmic one.
\subsection 4.3 Renormalization constants

In order to renormalize the correlation functions (4.12) -- (4.15), we 
first have to express the bare parameters $m_0$ and $g_0$ 
through the renormalized ones.
In a mass independent renormalization scheme, we have
\equation{
  g_R^2 = g_0^2 Z_g(g_0^2, a\mu)
\enum}
\equation{
  m_R = m_q Z_m(g_0^2, a\mu), \qquad m_q = m_0 - m_c
\enum}
We choose
\equation{
  m_R = 0
\enum}
To the order considered, the required substitution is then given by 
\equation{
  g_0^2 = g_R^2 + O(g_R^4) 
\enum}
With $m_R = 0$, we have
\equation{
  m_0 = m_c^{(1)}g_R^2 + O(g_R^4)
\enum}
For the value of $m_c^{(1)}$, we take the one computed in [\ref{sint}]
\equation{
  am_c^{(1)} = -0.4342856(3)
\enum}
We are now ready to extract the renormalization constants 
of the flavor singlet operators. We define the normalized correlation functions:
\equation{
h_{qq} = {f_{qq}\over \sqrt{f_1}}, \qquad
h_{qg} = {f_{qg}\over \sqrt{f_1}}
\enum}
\equation{
h_{gg} = {f_{gg}\over \sqrt{G_1}}, \qquad
h_{gq} = {f_{gq}\over \sqrt{G_1}}
\enum}
The relations between the operators and the renormalization
constants (4.2) and (4.3) for the correlation functions are
\equation{
h^{(R)}_{\alpha\beta} = \sum_{\gamma = q,g}Z_{\alpha\gamma}h_{\beta\gamma},\qquad \alpha,\beta=q,g
\enum}
The generic expansion of the $f$, $h$ and $Z$ functions is
\equation{
f = f^{(0)} + g_0^2 f^{(1)} + O(g_0^4)
\enum}
\equation{
h = h^{(0)} + g_0^2 h^{(1)} + O(g_0^4)
\enum}
\equation{
Z = Z^{(0)} + g_0^2 Z^{(1)} + O(g_0^4)
\enum}
For $h_{qg}$ or $h_{gq}$, $h^{(0)} = 0$. For $Z_{qq}$ or $Z_{gg}$, $Z^{(0)} = 1$, and for $Z_{qg}$ and $Z_{gq}$, $Z^{(0)} = 0$.
By expanding eqs.~(4.36) to order $O(g_R^2)$, we have
\equation{\eqalign{
h^{(R)}_{qq}(x_0,\theta) \ & = \ h_{qq}^{(0)} + g_R^2 \left\{h_{qq}^{(1)} + 
{\partial m_0\over \partial g_R^2} {\partial \over \partial m_0}
h_{qq}^{(0)} + Z^{(1)}_{qq}h_{qq}^{(0)}  \right\} + O(g_R^4) \cr
& = \ h_{qq}^{(0)} + g_R^2 \left\{h_{qq}^{(1)}  + 
m_c^{(1)} {\partial \over \partial m_0}
h_{qq}^{(0)} + Z^{(1)}_{qq} h_{qq}^{(0)} \right\} + O(g_R^4)
}
\enum}
\equation{
h^{(R)}_{gq}(x_0) = g_R^2 \left\{h_{gq}^{(1)} + Z^{(1)}_{qg}h_{gg}^{(0)}  \right\} + O(g_R^4)
\enum}
\equation{
h^{(R)}_{qg}(x_0,\theta) = g_R^2 \left\{h_{qg}^{(1)} + Z^{(1)}_{gq}h_{qq}^{(0)}  \right\} + O(g_R^4)
\enum}
\equation{
h^{(R)}_{gg}(x_0) = h_{gg}^{(0)} + g_R^2 \left\{h_{gg}^{(1)} + 
+ Z^{(1)}_{gg}h_{gg}^{(0)}  \right\} + O(g_R^4)
\enum}
where the amplitudes (4.40), (4.41) and (4.42)
on the right hand side are to be evaluated at $m_0 = 0$. In order to avoid technical problems 
without compromising the precision, we put $m_0 = 1.0 \times 10^{-10}$ in our programs.
\vskip 0.3cm
The renormalization conditions for the $h$'s read
\equation{
h^{(R)}_{qq} = h_{qq}^{(0)}, \qquad {\rm{with}} \quad x_0 = L/2, \quad  \theta_1 = 0.1, \quad  \mu = 1/L
\enum}
\equation{
h^{(R)}_{qg} = 0, \hbox{\hskip 1.13cm} {\rm{with}} \quad  x_0 = L/2, \quad  \theta_1 = 0.1, \quad  \mu = 1/L 
\enum}
\equation{
h^{(R)}_{gq} = 0, \hbox{\hskip 1.13cm} {\rm{with}} \quad  x_0 = L/2, \quad  \mu = 1/L
\enum}
\equation{
h^{(R)}_{gg} = h_{gg}^{(0)}, \qquad {\rm{with}} \quad  x_0 = L/2, \quad  \mu = 1/L
\enum}
Since the only dependence on the lattice spacing is through the combination
$a/L$, the continuum limit is equivalent to the limit $L/a\ \to \ \infty$, i.e. 
the number of points $N\ \to \ \infty$. By imposing the renormalization conditions
(4.44) -- (4.47), we obtain
\equation{
Z_{qq}^{(1)}\biggl(\theta_1, {x_0\over L},{a\over L}\biggr) = 
-{f_{qq}^{(1)}\over f_{qq}^{(0)}} - 
m_c^{(1)} {1\over f_{qq}^{(0)}}{\partial f_{qq}^{(0)} \over \partial m_0} +
{1\over 2}{f_1^{(1)}\over f_1^{(0)}}
\enum}
\equation{
Z_{qg}^{(1)}\biggl({x_0\over L},{a\over L}\biggr) = -{f_{gq}^{(1)}\over f_{gg}^{(0)}}
\enum}
\equation{
Z_{gq}^{(1)}\biggl(\theta_1,{x_0\over L},{a\over L}\biggr) = -{f_{qg}^{(1)}\over f_{qq}^{(0)}}
\enum}
\equation{
Z_{gg}^{(1)}\biggl({x_0\over L},{a\over L}\biggr) = -{f_{gg}^{(1)}\over f_{gg}^{(0)}} + 
{1\over 2}{G_1^{(1)}\over G_1^{(0)}}
\enum}
\section 5. Analysis and results

In order to separate the logarithmic coefficients from the finite terms in the renormalization 
constants (4.47) -- (4.50), and to get rid of the lattice artefacts embedded in their own definition, 
we apply a technique based on combinations of the Z's at different values of $N=L/a$
[\ref{pert}, \ref{numtec}]. From general arguments, it is known that the $N$ dependence including 
lattice artefacts can be parametrized as follows
\equation{
Z^{(1)}(N) = A + B\log(N) + \sum_{k=1}^\infty {1\over N^k} \{ C_k + D_k \log(N) \}, \qquad N = {L\over a}
\enum}
where all the coefficients of the expansion depend on the details of the calculation (boundary 
conditions, operator representations, etc.), except for $B$, which
is related to the leading anomalous dimension and is therefore scheme independent. 
In order to check that our $Z$'s reproduce the correct logarithmic coefficients, we introduce
 a numerical {\it logarithmic} derivative 
\equation{
\Delta^{(0)} (N) \ \equiv \ {N\over 2\eta}\ [Z^{(1)}(N + \eta) - Z^{(1)}(N - \eta)]
\enum}
with $\eta = 2$ for $Z^{(1)}_{qq}$ or $Z^{(1)}_{gq}$ and $\eta = 4$ for $Z^{(1)}_{qg}$ or $Z^{(1)}_{gg}$,
depending upon the presence of fermion or gluon sources, which are computed respectively on lattice
sizes multiple of two or four. The quantity $\Delta^{(0)}(N)$ slowly approaches $B$ with a rate 
proportional to $1/N$.  Infact, it can be expanded according to
\equation{
\Delta^{(0)}(N) = B + \tilde C_1 g^{(0)}(N) + \tilde D_1 h^{(0)}(N) + \tilde C_2 e^{(0)}(N) + \tilde D_2 m^{(0)}(N)\ + \ 
O\biggl({1\over N^3}\biggr)
\enum}
with new coefficients and the auxiliary functions
\equation{\eqalign{
& g^{(0)}(N) = {1\over N}, \quad \ \ h^{(0)}(N) = {1\over N} \log(N) \cr
& e^{(0)}(N) = {1\over N^2}, \quad m^{(0)}(N) = {1\over N^2} \log(N)
} \enum}
In order to have a safe continuum extrapolation of (5.3), terms of order $O(1/N)$ should be absent. 
This can be arranged with a simple procedure. The first step consists in building the quantity
\equation{
\Delta^{(1)}(N) \ \equiv \ {\ \Delta^{(0)}(N) g^{(0)}(N + \eta) - \Delta^{(0)}(N + \eta) g^{(0)}(N) \ \over g^{(0)}(N + \eta) - g^{(0)}(N)}
\enum}
which, by the same token, can be expanded as
\equation{
\Delta^{(1)}(N) = B + \hat C_1 g^{(1)}(N) + \hat C_2 h^{(1)}(N) + \hat D_2 e^{(1)}(N)\ + \  O\biggl({1\over N^3}\biggr)
\enum}
where a new set of coefficients and auxiliary functions
\equation{
g^{(1)}(N) = {1\over 2}\log\biggl(1 + {\eta \over N}\biggr);
\qquad h^{(1)}(N),e^{(1)}(N) \mathop{\longrightarrow}\limits^{N \ \to \ \infty} O\biggl({1\over N^2}\biggr)
\enum}
have been introduced. Note that, at this point, the only auxiliary function of order $O(1/N)$ is $g^{(1)}(N)$, and it can be removed with a second subtraction step, analogous to the previous one. By defining
\equation{
\Delta^{(2)}(N) \ \equiv \ {\ \Delta^{(1)}(N) g^{(1)}(N + \eta) - \Delta^{(1)}(N + \eta) g^{(1)}(N) \ 
\over g^{(1)}(N + \eta) - g^{(1)}(N)}
\enum}
one can easily convince himself that all the terms of order $O(1/N)$ have been removed, so that 
\equation{
\Delta^{(2)}(N) = B \ + \ O\biggl({1\over N^2}\biggr)
\enum}
and the logarithmic coefficient $B$ can be extracted with a good precision through an ordinary fit. Obviously, every subtraction
 step cancels out part of the result, and a big precision is required in order to avoid rounding 
effects. For this reason, all the computations have been done in double precision, as shown in Tables~1~--~3.
Numerical results for the anomalous dimensions are reported on second column of Table~4, and are compared with
 their analytic values, obtained from dimensional regularization. It has to be noted that $\gamma_{gg}^{(gl)} = 0$ 
and this is verified with a good precision by our computation. The results for $\gamma_{qq}$ are omitted, 
because they are the same ones as in the non singlet calculation, and can be found in [\ref{bucpal}].
\vskip 0.3cm
The extraction of the correct logarithmic coefficients shows the consistency of the calculation.  
The determination of the finite scheme dependent terms is obtained by subtracting the logarithmic
 divergences with the exact coefficients. This generates the {\it subtracted renormalization constants}
\equation{
Z^{(1)}_{\rm sub}(N) = A\ + \ \sum_{k=1}^\infty {1\over N^k} \{ C_k + D_k \log(N) \}
\enum}
that are again combined in order to suppress lattice artefacts.
Following eqs. (4.8) -- (4.11), we parametrize our continuum results as
\equation{
Z_{qq}(a/L)~=~1 + g_R^2\Big(\gamma_{qq} \log{a\over L} + B^{SF}_{qq}\Big)
\enum} 
\equation{
Z_{qg}(a/L)~=~ g_R^2 N_f\Big(\gamma_{qg} \log{a\over L} + B^{SF}_{qg}\Big)
\enum}
\equation{
Z_{gq}(a/L)~=~ g_R^2 \Big(\gamma_{gq} \log{a\over L} + B^{SF}_{gq}\Big)
\enum}
\equation{
Z_{gg}(a/L)~=~1 + g_R^2 \Big[N_f\Big(\gamma_{gg}^{(fe)} \log{a\over L} + [B^{SF}_{gg}]^{(fe)}\Big) 
+ \gamma_{gg}^{(gl)} \log{a\over L} + [B^{SF}_{gg}]^{(gl)}\Big]
\enum}
The values of the extrapolated $B_{\alpha\beta}^{SF}$ ($\alpha,\beta = q,g$) are reported on first column of Table~4. Again, $B_{qq}^{(SF)}$ is omitted, because it can be found in [\ref{bucpal}].
\vskip 0.4cm
The impact of lattice artefacts on the continuum approach of the finite coefficients $B_{\alpha\beta}^{SF}$ can be estimated dividing $Z_{\rm sub}^{(1)}$ by the continuum fit of the finite coefficients. We introduce the following {\it test} functions
\equation{
\sigma_{gg}^{(fe)}\biggl({L\over a}\biggr) = {[Z^{(1)}_{gg}]^{(fe)} - \gamma_{gg}^{(fe)} \log{a\over L} \over [B^{SF}_{gg}]^{(fe)}}
\enum}
\equation{
\sigma_{gg}^{(gl)}\biggl({L\over a}\biggr) = {[Z^{(1)}_{gg}]^{(gl)}\over \ \ [B^{SF}_{gg}]^{(gl)}\ \ }
\enum}
\equation{
\sigma_{gq}\biggl({L\over a}\biggr) = {Z^{(1)}_{gq} - \gamma_{gq} \log{a\over L}\over \ B^{SF}_{gq}\ }
\enum}
\equation{
\sigma_{qg}\biggl({L\over a}\biggr) = {Z^{(1)}_{qg} - \gamma_{qg} \log{a\over L}\over \ B^{SF}_{qg}\ }
\enum}
which converge to one in the continuum limit and differ from one because of the lattice artefacts.
Note that $Z^{(1)}$ and $Z_{\rm sub}^{(1)}$ coincide for the gluon--gluon correlation, because
the anomalous dimension is zero. In each of Figs.~13 -- 16 one curve refers to the 
unimproved case, and the other one shows the effect of the improvement procedure described above.  
After the suitable combinations, the continuum extrapolation is much more safe.
\vskip 0.4cm
The finite part of the renormalization constants can be used to match 
experimental results extracted at high energy in a specific continuum
scheme like $\overline{MS}$ and lattice results calculated nonperturbatively at
low energy and evolved to high energy within this particular lattice scheme (the SF scheme).
The matching coefficients are given by
\equation{
B_{\alpha\beta}^{\rm match} = B_{\alpha\beta}^{SF} - B_{\alpha\beta}^{\overline{MS}}, \qquad \alpha,\beta = q,g
\enum}
The values of $B_{\alpha\beta}^{\overline{MS}}$ can be found in [\ref{cost}] and the matching coefficients 
come out to be
\equation{
\left\{\matrix{  [B_{gg}^{\rm match}]^{(gl)} & = & -0.4697(1) \cr 
 	         [B_{gg}^{\rm match}]^{(fe)} & = & \ 0.1044(1) \cr 
		 B_{gq}^{\rm match}  & = & -0.04162(1) \cr 
		 B_{qg}^{\rm match}  & = & -0.1195(1)\ .  }\right .
\enum}
This is the result of the calculation.

\section 6. Conclusions

This computation put the premises for a lattice non perturbative calculation of
the amount of gluon in a hadron from first principles in the SF scheme.
The new definition of a gauge invariant gluon source
might also be used for a novel definition of $\alpha_s$, or for further studies of 
non perturbative aspects of the gluon propagation. Preliminary numerical simulations show
the feasibility of our definition [to appear...].
\vskip 2.0cm
{\bf Acknolwedgments}
\vskip 0.2cm
We are grateful to S. Sint and G. de Divitiis for discussions on possible definitions 
of gluon source operators in the SF scheme. We also express thanks to A. Bucarelli and V. D'Achille
for their collaboration in a preliminary stage of the calculation. 
\vfill\eject
\appendix A Strength tensor perturbative expansion

In this appendix we report the perturbative expansion of the lattice strength tensor $F_{\mu\nu}$
in time--momentum representation, up to order $g_0$. Fourier transforms are required only along the space directions
by the absence of periodic boundary conditions in time, which characterizes the SF scheme. This produces an asymmetry
between temporal and spatial Lorentz indices and makes the algebra a bit more
elaborate. The perturbative expansion is defined by the formula
\equation{
F_{\mu\nu}(x) \ = \ \sum_{k=0}^\infty \ g_0^k \ F_{\mu\nu}^{(k)}(x); \qquad \qquad F_{\mu\nu}^{(k)}(x) = - F_{\nu\mu}^{(k)}(x),  \qquad \forall k \in \Nset
\enum}
The $O(g_0^0)$ terms are given by
\equation{\eqalign{
F^{(0)}_{jk}(x) & = {i\over a} {1\over L^3} \sum_{\bp} \e^{i\bp \cdot \bx}
\left[\cos({ap_k\over 2}) \sin(ap_j) \tilde q^a_k(x_0;\bp) - \cos({ap_j\over 2}) \sin(ap_k) \tilde q^a_j(x_0;\bp) \right]T^a
}
\enum}
\equation{\eqalign{
F^{(0)}_{0k}(x) & = {1\over 2a} {1\over L^3} \sum_{\bp} \e^{i\bp \cdot \bx}
\biggl\{ \cos({ap_k\over 2}) [\tilde q^a_k(x_0 + a;\bp) - \tilde q^a_k(x_0 - a;\bp)]\ + \cr
& - \ i \sin(ap_k) [\tilde q^a_0(x_0;\bp) + \tilde q^a_0(x_0 - a;\bp)] \biggr\} T^a
}
\enum}
and the $O(g_0)$ ones are
\equation{\eqalign{
F^{(1)}_{jk}(x) & = {g_0\over 4} {1\over L^6} \sum_{\bp,\bq} \e^{i(\bp + \bq) \cdot \bx}
\left\{ 2 \left[ \cos({ap_j\over 2} + aq_j)\cos({aq_k\over 2}) + 
\cos({aq_k\over 2} + ap_k)\cos({ap_j\over 2}) \right. \right. + \cr
& - \left. \left.  \cos({ap_j\over 2})\cos({aq_k\over 2}) + 
\cos({ap_j\over 2} + aq_j)\cos({aq_k\over 2} + ap_k) \right] 
\tilde q^a_j(x_0;\bp) \tilde q^b_k(x_0;\bq) \right. + \cr
& + \left.  \sin({ap_j\over 2} + {aq_j\over 2}) 
[ \sin(aq_k) - \sin(ap_k)] \tilde q^a_j(x_0;\bp) \tilde q^b_j(x_0;\bq) \right. + \cr
& + \left.  \sin({ap_k\over 2} + {aq_k\over 2}) 
[ \sin(ap_j) - \sin(aq_j)] \tilde q^a_k(x_0;\bp) \tilde q^b_k(x_0;\bq)
\right\}
}
\enum}
\equation{\eqalign{
F^{(1)}_{0k}(x) & = {g_0\over 4} {1\over L^6} \sum_{\bp,\bq} \e^{i(\bp + \bq) \cdot \bx}
\left\{ \left[ \cos({aq_k\over 2} + ap_k) + \cos({aq_k\over 2}) \right] \right. \times \cr
& \times \left. \left[ \tilde q^a_0(x_0;\bp) \tilde q^b_k(x_0 + a;\bq) + 
\tilde q^a_0(x_0 - a;\bp) \tilde q^b_k(x_0 - a;\bq) \right] \right. + \cr
& + \left.  [\cos(ap_k + {aq_k\over 2}) - \cos({aq_k\over 2})]  \right. \times \cr
& \times \left. [ \tilde q^a_0(x_0;\bp) \tilde q^b_k(x_0;\bq) + 
\tilde q^a_0(x_0 - a;\bp) \tilde q^b_k(x_0;\bq)] \right. + \cr
& + \left.  i \sin(ap_k) [ \tilde q^a_0(x_0;\bp) \tilde q^b_0(x_0;\bq) - 
\tilde q^a_0(x_0 - a;\bp) \tilde q^b_0(x_0 - a;\bq)] \right. + \cr
& + \left.   i \sin({ap_k\over 2} + {aq_k\over 2}) 
[ \tilde q^a_k(x_0;\bp) \tilde q^b_k(x_0 + a;\bq) - 
\tilde q^a_k(x_0;\bp) \tilde q^b_k(x_0 - a;\bq)]
\right\} f^{abc}T^c
}
\enum}
\appendix B Feynman rules for the action

In this appendix we report the Feynman rules in the
time--momentum representation for the gauge part of the action, which never appeared
in the literature. The Feynman rules for the fermion one are given in [\ref{pert}].
The gauge part of the action is composed by a pure gauge term, a ghost term
and a measure term. The perturbative expansion in powers of $g_0$ reads
\equation{
S_G[q] = \sum_{k=0}^{\infty} g_0^k S_G^{(k)}[q]
\enum}
\equation{
S_m[q] = \sum_{k=1}^{\infty} g_0^{2k} S_m^{(2k)}[q]
\enum}
\equation{
S_{FP}[q,c,\bar c] = \sum_{k=0}^{\infty} g_0^k S_{FP}^{(k)}[q,c,\bar c]
\enum}
\noindent where every term in the sums of eqs.~(B.1) -- (B.3) has several
contributions in the time--momentum representation, due to the fact that we must 
separate spatial and temporal Lorentz indices.
The gauge action $S_G$ and the ghost action $S_{FP}$ are defined in
[\ref{pert}]. The measure term is defined in any standard lattice gauge theory book 
(see for example [\ref{montvay}]). 
\subsection B.1 Vertices with a non zero continuum limit

The time--momentum representation of the {\it three--gluons} vertex separates in three contributions
\equation{
S_G^{(1)}[q] = \sum_{i=a,b,c}\ S_G^{(1,i)}[q]
\enum}
with the explicit expressions (Feynman graphs are reported in Fig.~8)
\equation{\eqalign{
S_G^{(1,a)}[q] & = {1\over L^6} \sum_{\bl, \bm} a \sum_{x_0 = a}^{T-a} 
{i\over a} \cos\biggl[{a\over 2}(q_k - p_k)\biggr] \ \times \cr
& \times \cos\biggl[{a\over 2}(p_j + q_j)\biggr] 
\tilde q^a_k(x_0; -\bl-\bm) q^b_j(x_0; \bl) q^c_j(x_0; \bm) f^{abc}
}
\enum}
\equation{\eqalign{
S_G^{(1,b)}[q] & = {1\over L^6} \sum_{\bl,\bm} a \sum_{x_0 = a}^{T-a} 
{1\over a} \cos\biggl[{a\over 2}(p_j + q_j)\biggr] \ \times \cr
& \times \tilde q^a_0(x_0; -\bl-\bm) q^b_j(x_0; \bl) q^c_j(x_0+a; \bm) f^{abc}
}
\enum}
\equation{\eqalign{
S_G^{(1,c)}[q] & = {1\over L^6} \sum_{\bl,\bm} a \sum_{x_0 = a}^{T-a} 
\cos\biggl[{a\over 2}(q_k - p_k)\biggr] \times \cr
& \times {i\over 2a} \biggl[\tilde q^a_k(x_0 + a; -\bl-\bm)  + \tilde q^a_k(x_0; -\bl-\bm)\biggr]
q^b_0(x_0; \bl) q^c_0(x_0; \bm) f^{abc}
}
\enum}
\midinsert
\vbox{
\vskip0.0cm
\epsfxsize=10.0cm\hskip2.0cm\epsfbox{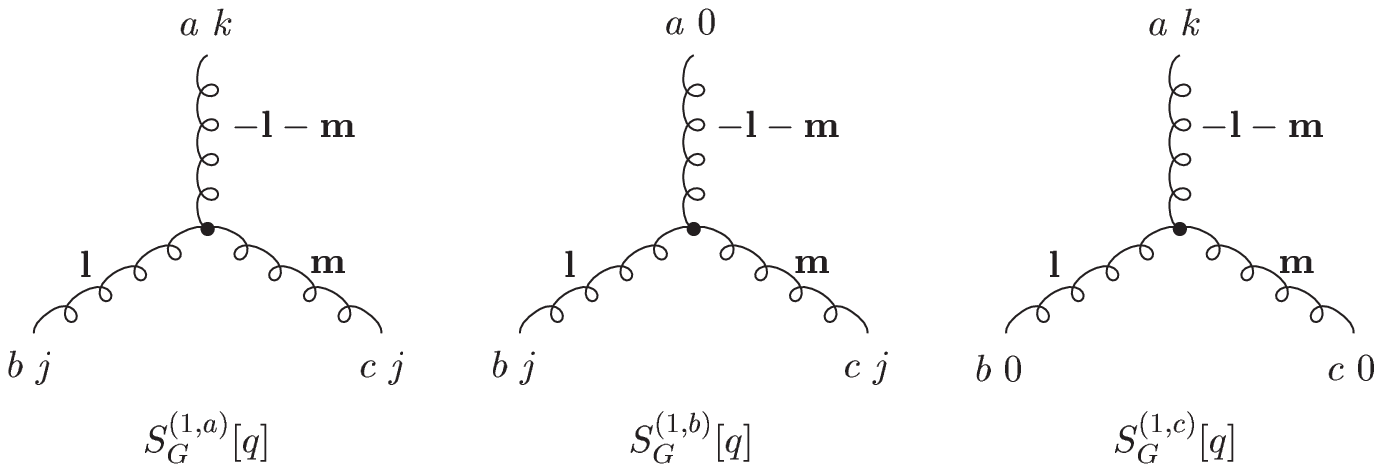}
\vskip0.2cm
\hskip 1.8cm \figurecaption{%
\eightrm Feynman diagrams of the three contributions to the {\it three--gluons} vertex.
}
\vskip0.0cm
}
\endinsert \vfill\eject
The time--momentum representation of the {\it four--gluon} has many contributions and we report
only the ones interested in computing our correlation functions. 
\equation{
S_G^{(2)}[q] = \sum_{i=a,b,c}\ S_G^{(2,i)}[q]
\enum}
with the explicit expressions (Feynman graphs are reported in Fig.~9)
\equation{\eqalign{
S_G^{(2,a)}[q] & = a^3 \sum_{\bx} a \sum_{x_0 = a}^{T-a} {1\over L^{12}} \sum_{\bl,\bm,\bn,\bp} 
\e^{i(\bl + \bm + \bn + \bp)\cdot\bx} \ \times \cr
& \times \biggl\{2 \tilde q_i^a (x_0;\bl) \tilde q_i^b (x_0;\bm) \tilde q_j^c (x_0;\bn) \tilde q_j^d (x_0;\bp) \biggr. \ \times \cr
& \times \biggl. \biggl[ \cos\bigl[{a\over 2}(l_j+m_j)\bigr] \cos\bigl[{a\over 2}(n_i+p_i)\bigr]
- \cos\bigl[{a\over 2}(l_i+m_i)\bigr] \cos\bigl[{a\over 2}(l_j-m_j)\bigr] \biggr. \biggr. \ + \cr
& - \biggl. \biggl. \cos\bigl[{a\over 2}(l_j+m_j)\bigr] \cos\bigl[{a\over 2}(p_i-n_i)\bigr] \biggr] 
-2 q_i^a (x_0;\bl) \tilde q_j^b (x_0;\bm) \tilde q_i^c (x_0;\bn) \tilde q_j^d (x_0;\bp) \biggr. \ \times \cr
& \times \biggl.  \cos\bigl[{a\over 2}(n_i-p_i)\bigr] \cos\bigl[{a\over 2}(m_j-l_j)\bigr] 
\biggr\} \ \tr(T^aT^bT^cT^d) 
}
\enum}
\equation{\eqalign{
S_G^{(2,b)}[q] & = a^3 \sum_{\bx} a \sum_{x_0 = a}^{T-a} {1\over L^{12}} \sum_{\bl,\bm,\bn,\bp}
\e^{i(\bl + \bm + \bn + \bp)\cdot \bx} \ \times \cr
& \times \biggl\{2 \tilde q_i^a (x_0;\bl) \tilde q_i^b (x_0;\bm) \tilde q_0^c (x_0;\bn) \tilde q_0^d (x_0;\bp) 
\sin\bigl({an_i\over 2}\bigr) \sin\bigl({ap_i\over 2}\bigr) \biggr. \ + \cr
& + \biggl. \tilde q_i^a (x_0 + a;\bl) \tilde q_i^b (x_0;\bm) \tilde q_0^c (x_0;\bn) \tilde q_0^d (x_0;\bp) 
\cos\bigl[{a\over 2}(n_i+p_i)\bigr] \biggr. \ + \cr
& + \biggl. \tilde q_i^a (x_0;\bl) \tilde q_i^b (x_0+a;\bm) \tilde q_0^c (x_0;\bn) \tilde q_0^d (x_0;\bp) 
\cos\bigl[{a\over 2}(n_i+p_i)\bigr] \biggr. \ + \cr
& - \biggl. \tilde q_i^a (x_0+a;\bl) \tilde q_i^b (x_0+a;\bm) \tilde q_0^c (x_0;\bn) \tilde q_0^d (x_0;\bp) 
\cos\bigl[{a\over 2}(n_i+p_i)\bigr] \biggr. + \cr
& + \biggl. \tilde q_0^a (x_0;\bn) \tilde q_i^b (x_0+a;\bl) \tilde q_i^c (x_0+a;\bm) \tilde q_0^d (x_0;\bp) 
\cos\bigl[{a\over 2}(n_i+p_i)\bigr] \biggr. \ + \cr
& - \biggl. 2 \tilde q_i^a (x_0;\bl) \tilde q_0^b (x_0;\bn) \tilde q_i^c (x_0+a;\bm) \tilde q_0^d (x_0;\bp) 
\cos\bigl[{a\over 2}(n_i+p_i)\bigr] \biggr\} \ \times \cr
& \times \tr(T^aT^bT^cT^d)
}
\enum}
\equation{\eqalign{
S_G^{(2,c)}[q] & = {a^3\over 24} \sum_{\bx} a \sum_{x_0 = a}^{T-a} {1\over L^{12}} \sum_{\bl,\bm,\bn,\bp}
\e^{i(\bl + \bm + \bn + \bp)\cdot \bx} \ \times \cr
& \times \biggl\{ \tilde q_i^a (x_0;\bl) \tilde q_i^b (x_0;\bm) \tilde q_i^c (x_0;\bn) \tilde q_i^d (x_0;\bp) \biggr. \ \times \cr
& \times \biggl. \biggl[ -10 + \sum_{k\neq i} \biggl( \cos(ap_k) + \cos(al_k) \biggr) - 12 
\sum_{k\neq i} \cos\bigl[a(l_k + m_k)\bigr] \biggr] \biggr. \ + \cr
& - \biggl. 2 \tilde q_i^a (x_0 + a;\bl) \tilde q_i^b (x_0 + a;\bm) \tilde q_i^c (x_0 + a;\bn) \tilde q_i^d (x_0 + a;\bp) 
\biggr. \ + \cr
& + \biggl. 8 \tilde q_i^a (x_0;\bl) \tilde q_i^b (x_0;\bm) \tilde q_i^c (x_0;\bn) \tilde q_i^d (x_0 + a;\bp) 
\biggr. \ + \cr
& + \biggl. 8 \tilde q_i^a (x_0;\bl) \tilde q_i^b (x_0 + a;\bm) \tilde q_i^c (x_0 + a;\bn) \tilde q_i^d (x_0 + a;\bp) 
\biggr. \ + \cr
& - \biggl. 12 \tilde q_i^a (x_0;\bl) \tilde q_i^b (x_0;\bm) \tilde q_i^c (x_0 + a;\bn) \tilde q_i^d (x_0 + a;\bp) 
\biggr\} \ \times \cr
& \times \tr(T^aT^bT^cT^d)
}
\enum}
\midinsert
\vbox{
\vskip0.0cm
\epsfxsize=10.0cm\hskip2.0cm\epsfbox{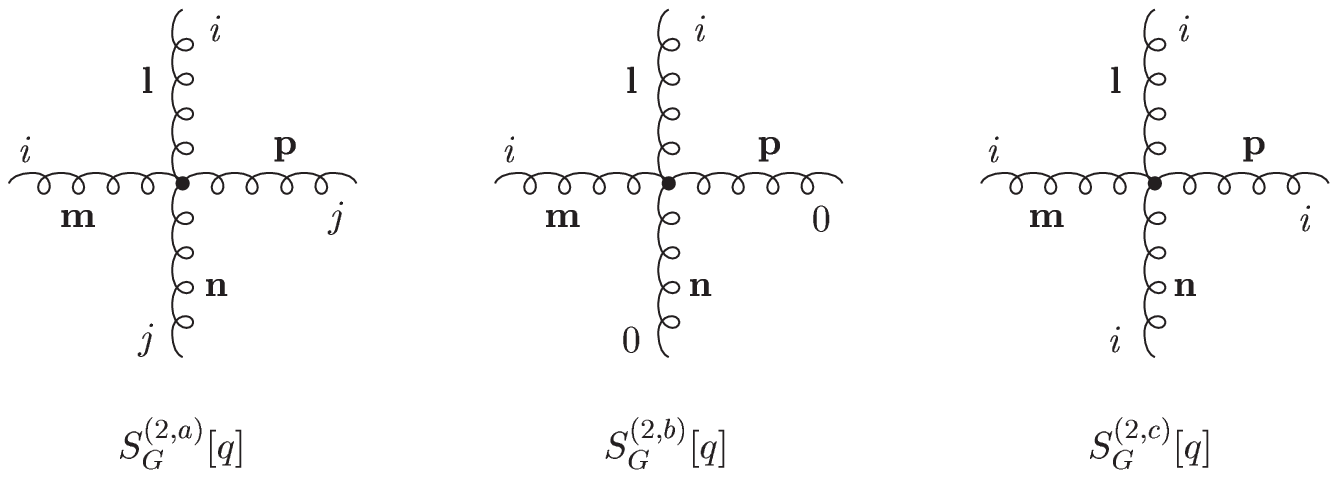}
\vskip0.2cm
\hskip 1.8cm \figurecaption{%
\eightrm Feynman diagrams of the three contributions to the {\it four--gluon} vertex.
}
\vskip0.0cm
}
\endinsert
Also the ghost term separates in two contributions (which are depicted in Fig.~10):
\equation{
S_{FP}^{(2)}[q,c,\bar c] = \sum_{i=a,b}\ S_{FP}^{(2,i)}[q,c,\bar c]
\enum}
and these are given by
\equation{\eqalign{
S_{FP}^{(1,a)}[q,c,\bar c] & = a \sum_{x_0 = a}^{T-a} {1\over L^6} \sum_{\bl,\bm} 
i \widehat{(l_k + m_k)} \cos\bigl( {a\over 2} m_k\bigr) 
\tilde{\bar c}^a (x_0;-\bl-\bm) \tilde q_k^b (x_0;\bl) \ \times \cr
& \times \tilde c^c (x_0;\bm) f^{abc}}
\enum}
\equation{\eqalign{
S_{FP}^{(1,b)}[q,c,\bar c] & = a \sum_{x_0 = a}^{T-a} {1\over L^6} \sum_{\bl,\bm} 
\tilde{\bar c}^a (x_0;-\bl-\bm) {1\over 2a} \biggl[\tilde q_0^b (x_0;\bl) \tilde c^c (x_0;\bm) \
+ \cr 
& + \tilde q_0^b (x_0;\bl) \tilde c^c (x_0 + a;\bm) - \tilde q_0^b (x_0-a;\bl) \tilde c^c (x_0 - a;\bm) \ + \cr
& - \tilde q_0^b (x_0-a;\bl) \tilde c^c (x_0;\bm)
\biggr] f^{abc}
}
\enum}
\midinsert
\vbox{
\vskip0.0cm
\epsfxsize=6.66cm\hskip3.3cm\epsfbox{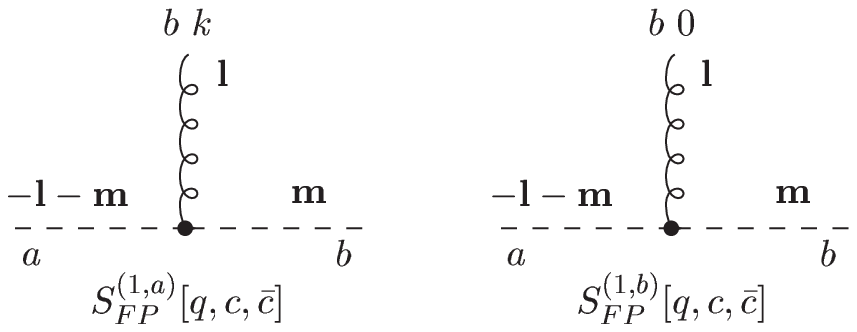}
\vskip0.2cm
\hskip 1.0cm \figurecaption{%
\eightrm Feynman diagrams of the two contributions to the {\it two-ghost} -- {\it one-gluon} vertex.
}
\vskip0.0cm
}
\endinsert
\subsection B.2 Vertices not existing in the continuum

The measure vertex is given by
\equation{
S_m^{(2)}[q] = {N_c\over 12 a^2} a \sum_{x_0 = a}^{T-a} {1\over L^3} \sum_{\bl}
\biggl[\sum_{k=1,2,3}\biggl(\tilde q_k^a (x_0;\bl) q_k^a (x_0;-\bl) \biggr) + 
q_0^a (x_0;\bl) q_0^a (x_0;-\bl) \biggr]
\enum}
\midinsert
\vbox{
\vskip0.0cm
\epsfxsize=0.9cm\hskip6.0cm\epsfbox{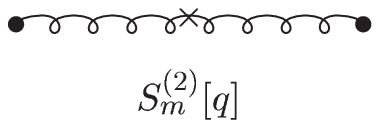}
\vskip0.2cm
\hskip 3.2cm \figurecaption{%
\eightrm Feynman diagram of the measure vertex.
}
\vskip0.0cm
}
\endinsert
The ghost action contributes also with a vertex which has a zero continuum limit. It is made by two
contributions (see Fig.~12):
\equation{
S_{FP}^{(2)}[q,c,\bar c] = \sum_{i=a,b}\ S_{FP}^{(2,i)}[q,c,\bar c]
\enum}
where
\equation{\eqalign{
S_{FP}^{(2,a)}[q,c,\bar c] & = {a^2\over 12} a \sum_{x_0 = a}^{T-a} {1\over L^9}\sum_{\bl,\bm,\bn}
\tilde{\bar c}^a (x_0;-\bl-\bm-\bn) \widehat{(l_k + m_k + n_k)} \hat{n}_k \times \cr
& \times \tilde q_k^b(x_0;\bl) \tilde q_k^c(x_0;\bm) \tilde c^d (x_0;\bn) f^{abe} f^{cde}
}
\enum}
\equation{\eqalign{
S_{FP}^{(2,b)}[q,c,\bar c] & = - {1\over 12} a \sum_{x_0 = a}^{T-a} {1\over L^9}\sum_{\bl,\bm,\bn}
\tilde{\bar c}^a (x_0;-\bl-\bm-\bn) \biggl[ q_0^b(x_0;\bl) \tilde q_0^c(x_0;\bm) \tilde c^d (x_0+a;\bn) \biggr. \ + \cr 
& - \biggl. q_0^b(x_0;\bl) \tilde q_0^c(x_0;\bm) \tilde c^d (x_0;\bn) 
- q_0^b(x_0-a;\bl) \tilde q_0^c(x_0-a;\bm) \tilde c^d (x_0;\bn) \biggr. \ + \cr
& + \biggl. q_0^b(x_0-a;\bl) \tilde q_0^c(x_0-a;\bm) \tilde c^d (x_0-a;\bn) \biggr] f^{abe} f^{cde}
}
\enum}
\midinsert
\vbox{
\vskip0.0cm
\epsfxsize=8.5cm\hskip2.5cm\epsfbox{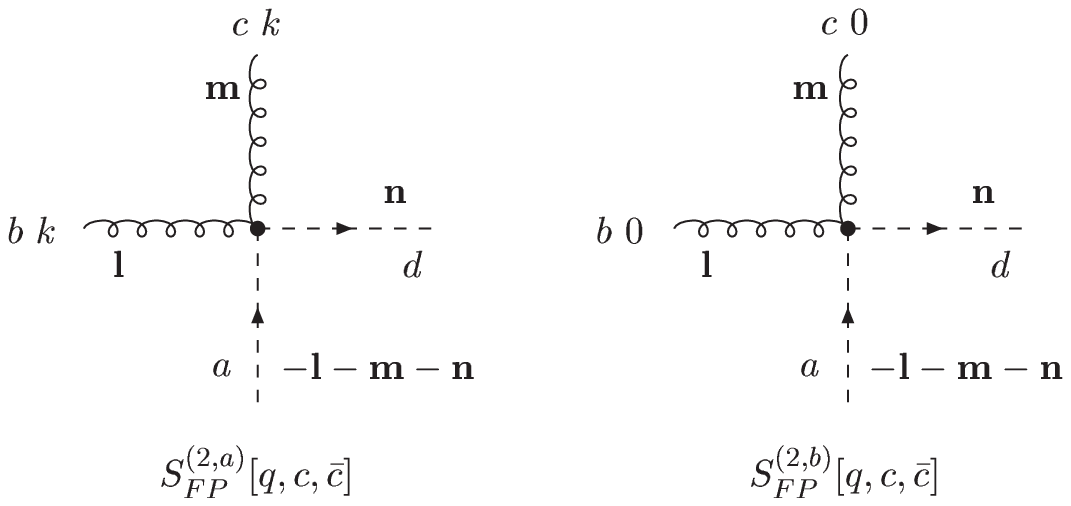}
\vskip0.2cm
\hskip 1.3cm \figurecaption{%
\eightrm Feynman diagrams of two contributions to the {\it two--ghost} -- {\it two--gluon} vertex.
}
\vskip0.0cm
}
\endinsert
\beginbibliography

\bibitem{strone}
M. G\"ockeler et al.,  \PRD D53 1996 2317-2325

\bibitem{strtwo}
C. Best et al., \PRD D56 1997 2743-2754

\bibitem{strthree}
D. Dolgov et al., \NPBproc 94 2001 303-306

\bibitem{strfour}
D. Dolgov et al., "Moments of Nucleon Light Cone Quark Distributions
Calculated in Full Lattice QCD" hep-lat/0201021

\bibitem{gjp}
M. Guagnelli et al., \NPB B542 1999 395-409

M. Guagnelli et al., \NPB B457 1999 153-156

M. Guagnelli et al., \NPB B459 1999 594-598

M. Guagnelli et al., \PLB B493 2000 77-81

\bibitem{gluonex}
J. Huston et al., \PRD D58 1998 114034

\bibitem{procrob} 
R. Petronzio, \NPBproc 83 2000 136-139


\bibitem{sffond}
M. L\" uscher, R. Narayanan, P. Weisz and U. Wolff, \NPB B384 1992 168

S.Sint, \NPB B421 1994 135

\bibitem{pert}
M. L\" uscher and P. Weisz, \NPB B479 1996 429

\bibitem{grone}
M. Baake, B. Gem\"unden and R.Oedingen, J.~Math.~Phys.~{\bf 23} (1982) 944 

J.E.Mandula, G.Zweig and J.Govaerts, \NPB B228 1983 91
 
\bibitem{grtwo}
G. Beccarini, M. Bianchi, S. Capitani and G. Rossi, \NPB B456 1995 271

M.G\"ockeler \etal, \PRD D54 1996 5705 

\bibitem{shitalk}
A. Shindler,  \NPBproc 83 2000 253-255

\bibitem{bucpal} 
A. Bucarelli, F. Palombi, R. Petronzio and A. Shindler, \NPB B552 1999 379

\bibitem{gross} 
D.J.Gross e F.Wilczeck, \PR D8 1973 3633 

D.J.Gross and F.Wilczeck, \PR D9 1974 980

\bibitem{sint} 
S. Sint, private notes (1996)


\bibitem{cost} 
S. Capitani, G. Rossi \NPBproc 53 1997 801-803

\bibitem{montvay}
I. Montvay, G. M\"unster, {\it{Quantum fields on a
lattice}} (Cambridge University Press)

\bibitem{numtec} 
M. L\"uscher, P. Weisz, \NPB B266 1986 309

\endbibliography
\vfill\eject
{\ninepoint
\topinsert
\newdimen\digitwidth
\setbox0=\hbox{\rm 0}
\digitwidth=\wd0
\catcode`@=\active
\def@{\kern\digitwidth}
\tablecaption{Ratio \ $G_1^{(1)} / G_1^{(0)}$}
\vskip 1.5ex
$$\vbox{\settabs\+x&xxxxxx&xxxxxxx&xxxxxxxxxxxxxxxxx&xxxxxxxxxx&xxxxxxxxxxxxxxxxxxx&xxxxx\cr
\thicktablerule
\vskip1ex
                \+& \hfill $T/a$ \hfill
                 && \hfill $R_1^{(gl)}$\hfill
                 && \hfill $R_1^{(fe)}$\hfill
                 &&  \cr
\vskip1.0ex
\thintablerule
\vskip1.5ex
  \+& \hfill $4$ \hfill
  &&  \hfill $-2.35241571640\times 10^0$ \hfill
  &&  \hfill $5.43688719630\times 10^{-1}$\hfill 
  &\cr
  \+& \hfill $8$ \hfill
  &&  \hfill $-2.89616691565\times 10^0$ \hfill
  &&  \hfill $2.91628523582\times 10^{-1}$ \hfill
  &\cr
  \+& \hfill $12$ \hfill
  &&  \hfill $-3.78563617887\times 10^0$ \hfill
  &&  \hfill $2.40540867861\times 10^{-1}$ \hfill 
  &\cr
  \+& \hfill $16$ \hfill
  &&  \hfill $-4.77638846812\times 10^0$ \hfill
  &&  \hfill $2.18562655437\times 10^{-1}$ \hfill 
  &\cr
  \+& \hfill $20$ \hfill
  &&  \hfill $-5.81427883554\times 10^0$ \hfill
  &&  \hfill $2.05144176593\times 10^{-1}$ \hfill
  &\cr
  \+& \hfill $24$ \hfill
  &&  \hfill $-6.87867683434\times 10^0$ \hfill
  &&  \hfill $1.95721360604\times 10^{-1}$ \hfill
  &\cr
  \+& \hfill $28$ \hfill
  &&  \hfill $-7.95977357383\times 10^0$ \hfill
  &&  \hfill $1.88582311080\times 10^{-1}$ \hfill
  &\cr
  \+& \hfill $32$ \hfill
  &&  \hfill $-9.05223296007\times 10^0$ \hfill
  &&  \hfill $1.82902372508\times 10^{-1}$ \hfill
  &\cr
  \+& \hfill $36$ \hfill
  &&  \hfill $-1.01528674964\times 10^1$ \hfill
  &&  \hfill $1.78225084856\times 10^{-1}$ \hfill
  &\cr
  \+& \hfill $40$ \hfill
  &&  \hfill $-1.12596369563\times 10^1$ \hfill
  &&  \hfill $1.74273581484\times 10^{-1}$ \hfill
  &\cr
  \+& \hfill $44$ \hfill
  &&  \hfill $-1.23711642225\times 10^1$ \hfill
  &&  \hfill $1.70868479333\times 10^{-1}$ \hfill
  &\cr
  \+& \hfill $48$ \hfill
  &&  \hfill $-1.34864798286\times 10^1$ \hfill
  &&  \hfill $1.67887576085\times 10^{-1}$ \hfill
  &\cr
  \+& \hfill $52$ \hfill
  &&  \hfill $-1.46048775931\times 10^1$ \hfill
  &&  \hfill $1.65244260811\times 10^{-1}$ \hfill
  &\cr
  \+& \hfill $56$ \hfill
  &&  \hfill $-1.57258283938\times 10^1$ \hfill
  &&  \hfill $1.62875144474\times 10^{-1}$ \hfill
  &\cr
  \+& \hfill $60$ \hfill
  &&  \hfill $-1.68489262682\times 10^1$ \hfill
  &&  \hfill $1.60732582532\times 10^{-1}$ \hfill
  &\cr
\vskip1ex
\thicktablerule
}$$
\endinsert}

{\ninepoint
\topinsert
\newdimen\digitwidth
\setbox0=\hbox{\rm 0}
\digitwidth=\wd0
\catcode`@=\active
\def@{\kern\digitwidth}
\tablecaption{Ratios $f_{qg}^{(1)} / f_{qq}^{(0)}$ and $f_{gq}^{(1)} / f_{gg}^{(0)}$}
\vskip 1.5ex
$$\vbox{\settabs\+x&xxxxxx&xxxxxxx&xxxxxxxxxxxxxxxxx&xxxxxxxxxx&xxxxxxxxxxxxxxxxxxx&xxxxx\cr
\thicktablerule
\vskip1ex
                \+& \hfill $T/a$ \hfill
                 && \hfill $f_{qg}^{(1)} / f_{qq}^{(0)}$\hfill
                 && \hfill $R_{gq}^{(fe)}$\hfill
                 &&  \cr
\vskip1.0ex
\thintablerule
\vskip1.5ex
  \+& \hfill $4$ \hfill
  &&  \hfill $-3.90445955289\times 10^{-2}$ \hfill
  &&  \hfill $1.76632912708\times 10^{-1}$\hfill 
  &\cr
  \+& \hfill $6$ \hfill
  &&  \hfill $-5.61960383511\times 10^{-2}$ \hfill
  &&  \hfill ---\hfill 
  &\cr
  \+& \hfill $8$ \hfill
  &&  \hfill $-6.84291322541\times 10^{-2}$ \hfill
  &&  \hfill $1.29586992658\times 10^{-1}$ \hfill
  &\cr
  \+& \hfill $10$ \hfill
  &&  \hfill $-7.81443039581\times 10^{-2}$ \hfill
  &&  \hfill ---\hfill 
  &\cr
  \+& \hfill $12$ \hfill
  &&  \hfill $-8.61855815840\times 10^{-2}$ \hfill
  &&  \hfill $1.15745211197\times 10^{-1}$ \hfill 
  &\cr
  \+& \hfill $14$ \hfill
  &&  \hfill $-9.30301253579\times 10^{-2}$ \hfill
  &&  \hfill ---\hfill 
  &\cr
  \+& \hfill $16$ \hfill
  &&  \hfill $-9.89813896977\times 10^{-2}$ \hfill
  &&  \hfill $1.08336211173\times 10^{-1}$ \hfill 
  &\cr
  \+& \hfill $18$ \hfill
  &&  \hfill $-1.04242788232\times 10^{-1}$ \hfill
  &&  \hfill ---\hfill 
  &\cr
  \+& \hfill $20$ \hfill
  &&  \hfill $-1.08956354293\times 10^{-1}$ \hfill
  &&  \hfill $1.03478207253\times 10^{-1}$ \hfill
  &\cr
  \+& \hfill $22$ \hfill
  &&  \hfill $-1.13224779604\times 10^{-1}$ \hfill
  &&  \hfill ---\hfill 
  &\cr
  \+& \hfill $24$ \hfill
  &&  \hfill $-1.17124551957\times 10^{-1}$ \hfill
  &&  \hfill $9.99540929596\times 10^{-2}$ \hfill
  &\cr
  \+& \hfill $26$ \hfill
  &&  \hfill $-1.20714100952\times 10^{-1}$ \hfill
  &&  \hfill ---\hfill 
  &\cr
  \+& \hfill $28$ \hfill
  &&  \hfill $-1.24039035245\times 10^{-1}$ \hfill
  &&  \hfill $9.72321408957\times 10^{-2}$ \hfill
  &\cr
  \+& \hfill $30$ \hfill
  &&  \hfill $-1.27135621901\times 10^{-1}$ \hfill
  &&  \hfill ---\hfill 
  &\cr
  \+& \hfill $32$ \hfill
  &&  \hfill $-1.30033166208\times 10^{-1}$ \hfill
  &&  \hfill $9.50373471243\times 10^{-2}$ \hfill
  &\cr
  \+& \hfill $34$ \hfill
  &&  \hfill $-1.32755681834\times 10^{-1}$ \hfill
  &&  \hfill ---\hfill 
  &\cr
  \+& \hfill $36$ \hfill
  &&  \hfill $-1.35323089945\times 10^{-1}$ \hfill
  &&  \hfill $9.32112990034\times 10^{-2}$ \hfill
  &\cr
  \+& \hfill $38$ \hfill
  &&  \hfill $-1.37752097648\times 10^{-1}$ \hfill
  &&  \hfill ---\hfill 
  &\cr
  \+& \hfill $40$ \hfill
  &&  \hfill $-1.40056853184\times 10^{-1}$ \hfill
  &&  \hfill $9.16554753604\times 10^{-2}$ \hfill
  &\cr
  \+& \hfill $42$ \hfill
  &&  \hfill $-1.42249442526\times 10^{-1}$ \hfill
  &&  \hfill ---\hfill 
  &\cr
  \+& \hfill $44$ \hfill
  &&  \hfill $-1.44340271231\times 10^{-1}$ \hfill
  &&  \hfill $9.03049553605\times 10^{-2}$ \hfill
  &\cr
  \+& \hfill $46$ \hfill
  &&  \hfill $-1.46338361917\times 10^{-1}$ \hfill
  &&  \hfill ---\hfill 
  &\cr
  \+& \hfill $48$ \hfill
  &&  \hfill $-1.48251588765\times 10^{-1}$ \hfill
  &&  \hfill $8.91149749248\times 10^{-2}$ \hfill
  &\cr
  \+& \hfill $50$ \hfill
  &&  \hfill $-1.50086864377\times 10^{-1}$ \hfill
  &&  \hfill ---\hfill 
  &\cr
  \+& \hfill $52$ \hfill
  &&  \hfill $-1.51850290135\times 10^{-1}$ \hfill
  &&  \hfill $8.80535191120\times 10^{-2}$ \hfill
  &\cr
  \+& \hfill $54$ \hfill
  &&  \hfill $-1.53547278304\times 10^{-1}$ \hfill
  &&  \hfill ---\hfill 
  &\cr
  \+& \hfill $56$ \hfill
  &&  \hfill $-1.55182651997\times 10^{-1}$ \hfill
  &&  \hfill $8.70969908092\times 10^{-2}$ \hfill
  &\cr
  \+& \hfill $58$ \hfill
  &&  \hfill $-1.56760727661\times 10^{-1}$ \hfill
  &&  \hfill ---\hfill 
  &\cr
  \+& \hfill $60$ \hfill
  &&  \hfill $-1.58285383623\times 10^{-1}$ \hfill
  &&  \hfill $8.62275524536\times 10^{-2}$ \hfill
  &\cr
\vskip1ex
\thicktablerule
}$$
\endinsert}

{\ninepoint
\topinsert
\newdimen\digitwidth
\setbox0=\hbox{\rm 0}
\digitwidth=\wd0
\catcode`@=\active
\def@{\kern\digitwidth}
\tablecaption{Ratio $f_{gg}^{(1)} / f_{gg}^{(0)}$}
\vskip 1.5ex
$$\vbox{\settabs\+x&xxxxxx&xxxxxxx&xxxxxxxxxxxxxxxxx&xxxxxxxxxx&xxxxxxxxxxxxxxxxxxxxxx&xxxxx\cr
\thicktablerule
\vskip1ex
                \+& \hfill $T/a$ \hfill
                 && \hfill $R_{gg}^{(gl)}$\hfill
                 && \hfill $R_{gg}^{(fe)}$\hfill
                 &&  \cr
\vskip1.0ex
\thintablerule
\vskip1.5ex
  \+& \hfill $4$ \hfill
  &&  \hfill $-7.42272842206\times 10^{-1}$ \hfill
  &&  \hfill $\ \ 4.32567853743\times 10^{-2}$\hfill 
  &\cr
  \+& \hfill $8$ \hfill
  &&  \hfill $-1.03655972167\times 10^0$ \hfill
  &&  \hfill $-1.42248562169\times 10^{-2}$ \hfill
  &\cr
  \+& \hfill $12$ \hfill
  &&  \hfill $-1.47142280292\times 10^0$ \hfill
  &&  \hfill $-3.03291610705\times 10^{-2}$ \hfill 
  &\cr
  \+& \hfill $16$ \hfill
  &&  \hfill $-1.95917324599\times 10^0$ \hfill
  &&  \hfill $-3.89219901277\times 10^{-2}$ \hfill 
  &\cr
  \+& \hfill $20$ \hfill
  &&  \hfill $-2.47274884132\times 10^0$ \hfill
  &&  \hfill $-4.48470953718\times 10^{-2}$ \hfill
  &\cr
  \+& \hfill $24$ \hfill
  &&  \hfill $-3.00105188219\times 10^0$ \hfill
  &&  \hfill $-4.93690136333\times 10^{-2}$ \hfill
  &\cr
  \+& \hfill $28$ \hfill
  &&  \hfill $-3.53866677061\times 10^0$ \hfill
  &&  \hfill $-5.30184104212\times 10^{-2}$ \hfill
  &\cr
  \+& \hfill $32$ \hfill
  &&  \hfill $-4.08261542990\times 10^0$ \hfill
  &&  \hfill $-5.60724863587\times 10^{-2}$ \hfill
  &\cr
  \+& \hfill $36$ \hfill
  &&  \hfill $-4.63111119327\times 10^0$ \hfill
  &&  \hfill $-5.86948398012\times 10^{-2}$ \hfill
  &\cr
  \+& \hfill $40$ \hfill
  &&  \hfill $-5.18300914136\times 10^0$ \hfill
  &&  \hfill $-6.09901327477\times 10^{-2}$ \hfill
  &\cr
  \+& \hfill $44$ \hfill
  &&  \hfill $-5.73753689001\times 10^0$ \hfill
  &&  \hfill $-6.30293363500\times 10^{-2}$ \hfill
  &\cr
  \+& \hfill $48$ \hfill
  &&  \hfill $-6.29415150120\times 10^0$ \hfill
  &&  \hfill $-6.48627573963\times 10^{-2}$ \hfill
  &\cr
  \+& \hfill $52$ \hfill
  &&  \hfill $-6.85245828489\times 10^0$ \hfill
  &&  \hfill $-6.65273259871\times 10^{-2}$ \hfill
  &\cr
  \+& \hfill $56$ \hfill
  &&  \hfill $-7.41216219458\times 10^0$ \hfill
  &&  \hfill $-6.80509342931\times 10^{-2}$ \hfill
  &\cr
  \+& \hfill $60$ \hfill
  &&  \hfill $-7.97303741242\times 10^0$ \hfill
  &&  \hfill $-6.94551510343\times 10^{-2}$ \hfill
  &\cr
\vskip1ex
\thicktablerule
}$$
\endinsert}

{\ninepoint
\topinsert
\newdimen\digitwidth
\setbox0=\hbox{\rm 0}
\digitwidth=\wd0
\catcode`@=\active
\def@{\kern\digitwidth}
\tablecaption{Constant and logarithmic coefficients as obtained numerically}
\vskip 1.5ex
$$\vbox{\settabs\+x&xxxxxxxxx&x&x&xxxxxxxxxxxx&xx&xxxxxxx&xxxxxxx&xx&xxxxxxxxxxxxxxxxxxxxxxxxxxxxxxx&xxxxx\cr
\thicktablerule
\vskip1.5ex
  \+& \hfill $[B^{SF}_{gg}]^{(gl)} \hskip 0.08cm=$ \hfill
  && \hfill $-0.4613(1)$\hfill
  && \hfill $|\gamma_{gg}^{(gl)}| <$\hfill
  && \hfill $\hskip 0.27cm 5.0\times 10^{-5}$
  && \hfill $$
  &&  \cr
\vskip1.5ex
\thintablerule
\vskip1.5ex
  \+& \hfill $[B^{SF}_{gg}]^{(fe)} =$ \hfill
  && \hfill $\hskip 0.27cm 0.1114(1)$\hfill
  && \hfill $\hskip 0.1cm\gamma_{gg}^{(fe)} =$\hfill
  && \hfill $-0.0083(1)N_f$
  && \hfill $(\ -{1\over 12\pi^2}N_f = -0.0084434...N_f\ )$
  &&  \cr
\vskip1.5ex
\thintablerule
\vskip1.5ex
  \+& \hfill $B^{SF}_{qg} \hskip 0.72cm=$ \hfill
  &&  \hfill $-0.1167(1)$ \hfill
  &&  \hfill $\hskip 0.1cm\gamma_{qg}\hskip 0.24cm =$\hfill 
  && \hfill $-0.0084(1)N_f$
  && \hfill $(\ -{1\over 12\pi^2}N_f = -0.0084434...N_f\ )$
  &\cr
\vskip1.5ex
\thintablerule
\vskip1.5ex
  \+& \hfill $B^{SF}_{gq} \hskip 0.72cm=$ \hfill
  &&  \hfill $-0.02614(1)$ \hfill
  &&  \hfill $\hskip 0.1cm\gamma_{gq}\hskip 0.24cm =$\hfill 
  &&  \hfill $-0.04502(1)$
  &&  \hfill $(\ -{4\over 9\pi^2} = -.045031...\ )$
  &\cr
\vskip1.5ex
\thicktablerule
}$$
\endinsert}

\vfill\eject

\ninepoint
\topinsert
\vbox{
\vskip0.0cm
\epsfxsize=7.5cm\hskip2.5cm\epsfbox{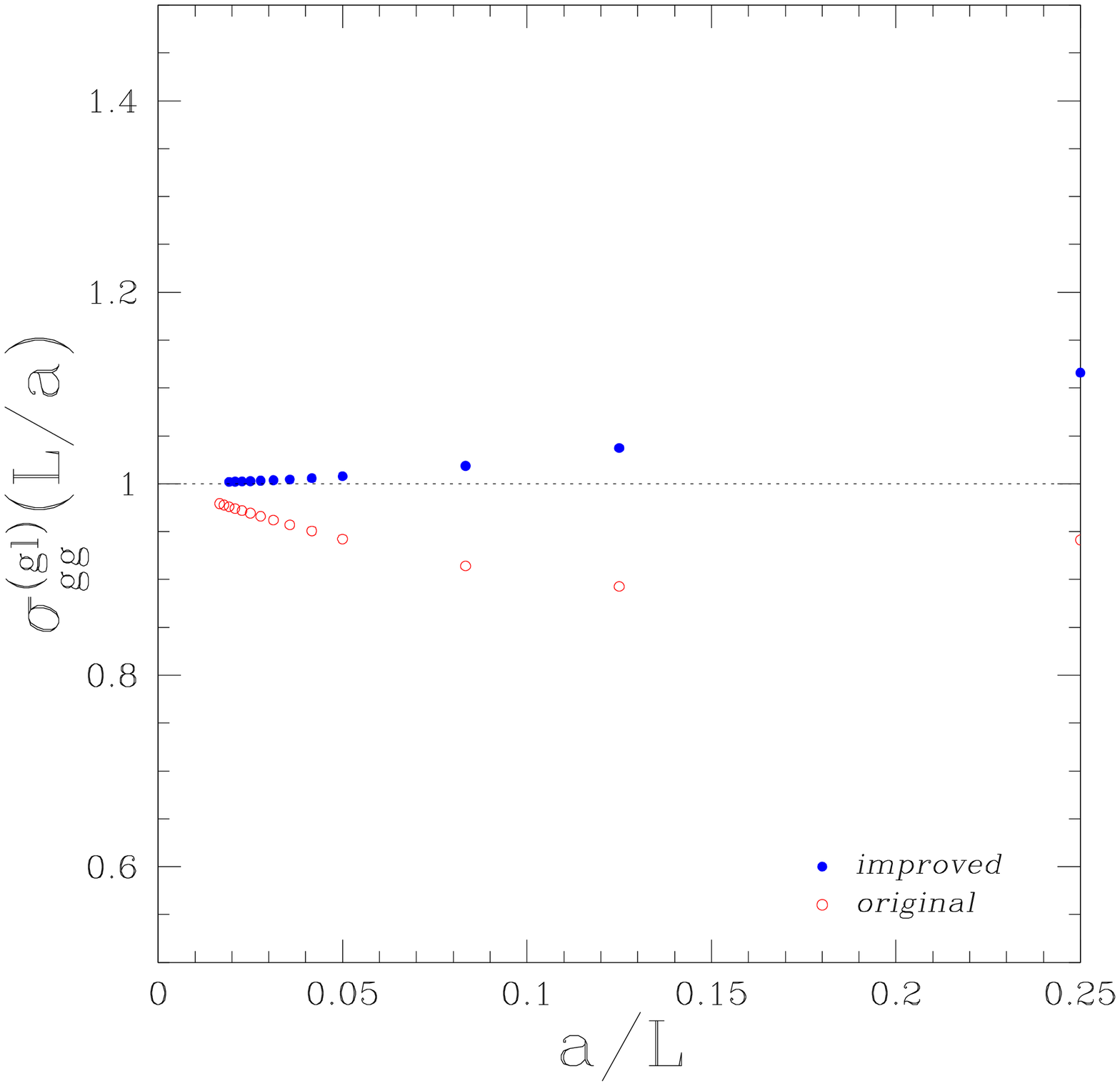}
\vskip0.2cm
\figurecaption{%
\eightrm Continuum approach of $[Z^{(1)}_{gg}]^{(gl)}$. Empty dots represent the original data, while the filled ones result from the cleaning procedure.
\eightrm 
}
\vskip0.0cm
}
\endinsert

\midinsert
\vbox{
\vskip0.0cm
\epsfxsize=7.5cm\hskip2.5cm\epsfbox{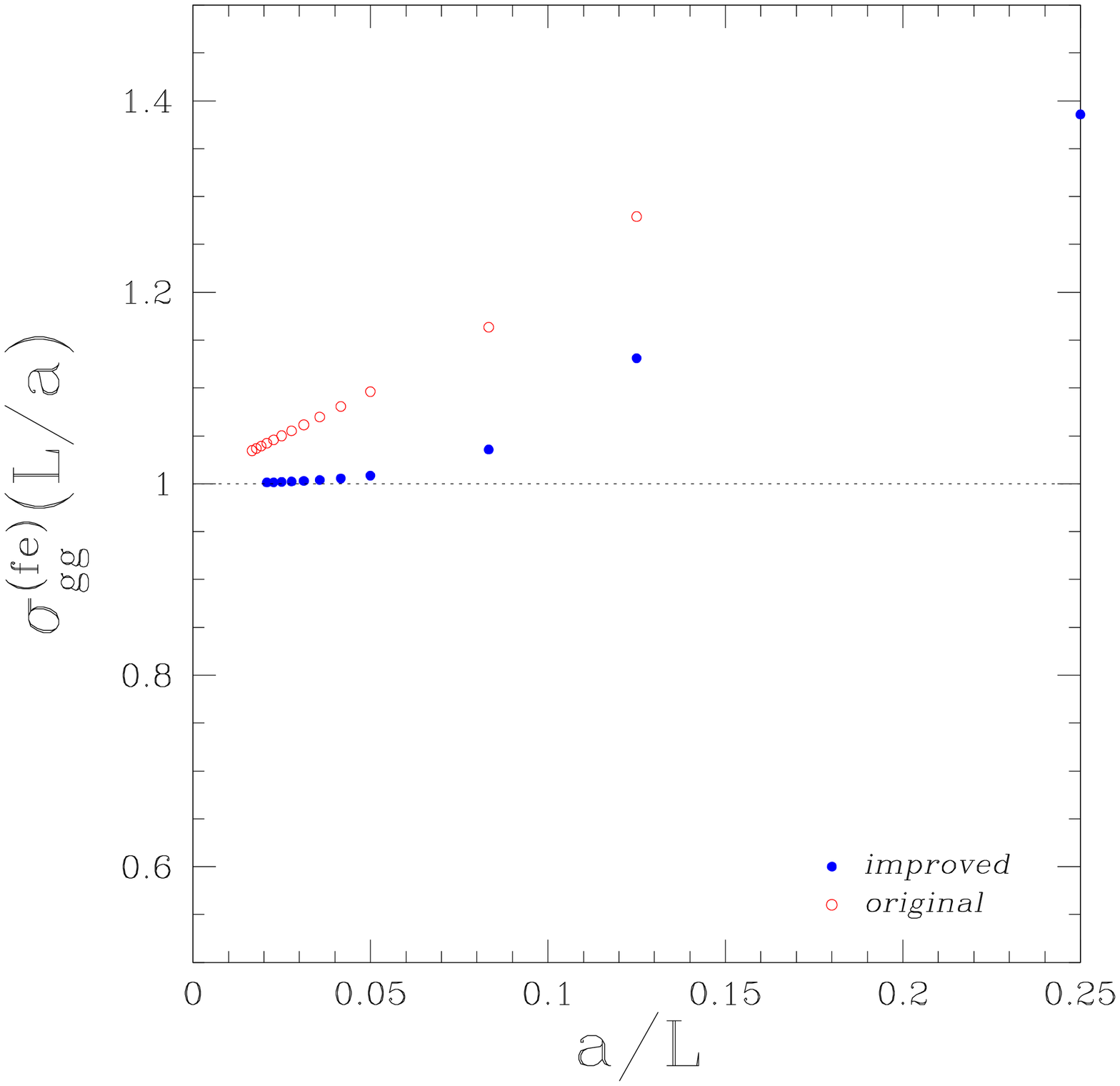}
\vskip0.2cm
\figurecaption{%
\eightrm Continuum approach of $[Z^{(1)}_{gg}]^{(fe)}$. Empty dots represent the original data, while the filled ones result from the cleaning procedure.
}
\vskip0.0cm
}
\endinsert

\topinsert
\vbox{
\vskip0.0cm
\epsfxsize=7.5cm\hskip2.5cm\epsfbox{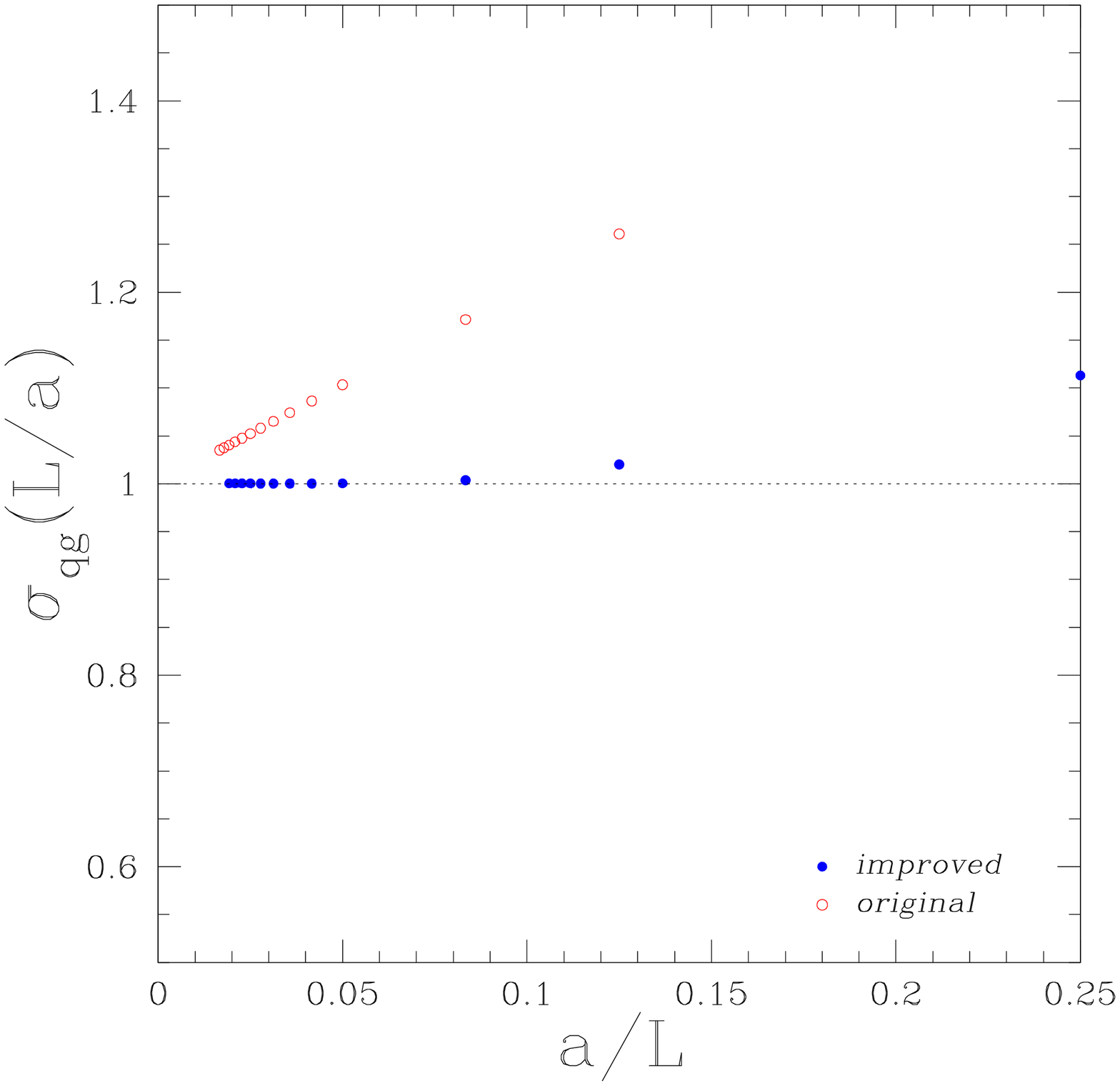}
\vskip0.2cm
\figurecaption{%
\eightrm Continuum approach of $Z^{(1)}_{qg}$. Empty dots represent the original data, while the filled ones result from the cleaning procedure.
}
\vskip0.0cm
}
\endinsert

\midinsert
\vbox{
\vskip0.0cm
\epsfxsize=7.5cm\hskip2.5cm\epsfbox{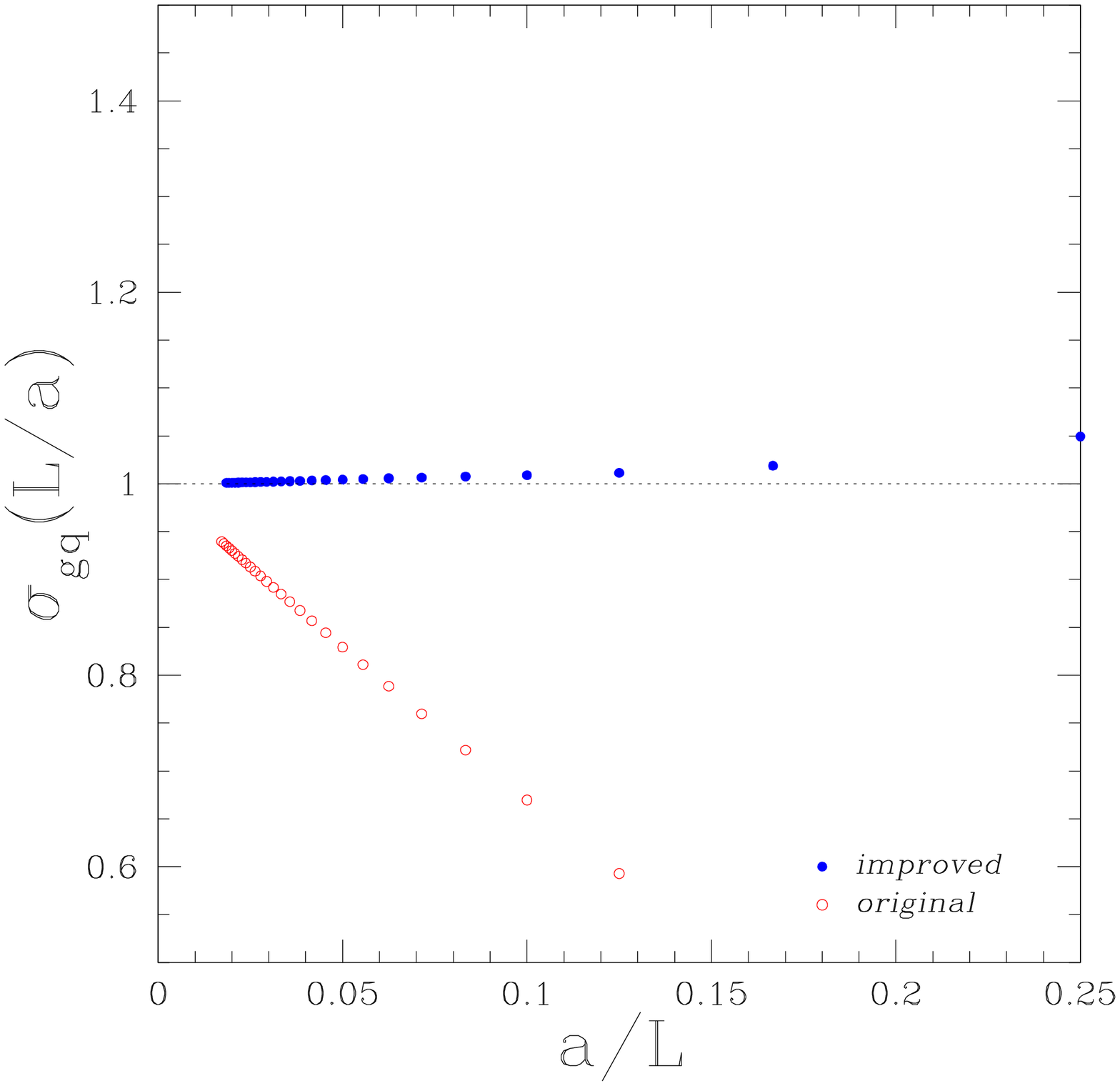}
\vskip0.2cm
\figurecaption{%
\eightrm Continuum approach of $Z^{(1)}_{gq}$. Empty dots represent the original data, while the filled ones result from the cleaning procedure.
}
\vskip0.0cm
}
\endinsert

\bye